\documentclass[aps,prd,superscriptaddress,amsmath,groupedaddress,twocolumn,nofootinbib,preprintnumbers]{revtex4}

\usepackage[colorlinks=true]{hyperref}
\usepackage{url}
\pdfoutput=1
\usepackage{ifpdf}

\usepackage{color,hhline,psfrag,rotating,amssymb}
\usepackage[hang,nooneline]{subfigure}
\usepackage{dcolumn}
\usepackage{stmaryrd}
\usepackage{bm}

\begin{document}
\title{New methods for $B$ meson decay constants and form factors from 
lattice NRQCD} 
\author{C. Hughes}
\email[]{chughes@fnal.gov}
\affiliation{Fermi National Accelerator Laboratory, Batavia, IL 60510, USA}
\affiliation{Department of Applied Mathematics and Theoretical Physics, University of Cambridge, Cambridge, CB3 0WA, UK}
\author{C. T. H. Davies}
\email[]{christine.davies@glasgow.ac.uk}
\affiliation{SUPA, School of Physics and Astronomy, University of Glasgow, Glasgow, G12 8QQ, UK}
\author{C. J. Monahan}
\affiliation{Institute for Nuclear Theory, University of Washington, Seattle, Washington 98195, USA}
\affiliation{New High Energy Theory Center and Department of Physics and Astronomy, Rutgers, the State University of New Jersey, 136 Frelinghuysen Road, Piscataway, NJ 08854, USA}
\collaboration{HPQCD collaboration}
\homepage{http://www.physics.gla.ac.uk/HPQCD}
\noaffiliation

\date{\today}

\preprint{FERMILAB-PUB-17-547-T}
\begin{abstract}
We determine the normalisation of scalar and pseudoscalar current  
operators made from non-relativistic $b$ quarks 
and Highly Improved Staggered light quarks in lattice Quantum 
Chromodynamics (QCD) through 
$\mathcal{O}(\alpha_s)$ and $\Lambda_{\text{QCD}}/m_b$. We use matrix elements 
of these operators to extract $B$ meson decay constants and form 
factors, then compare to those obtained using the standard 
vector and axial-vector operators. This provides a test of 
systematic errors in the lattice QCD determination of 
the $B$ meson decay constants and form factors. 
We provide a new value for the $B$ and $B_s$ meson decay constants 
from lattice QCD calculations on ensembles that include 
$u$, $d$, $s$ and $c$ quarks in the sea and those which have the $u/d$ quark 
mass going down to its physical value. 
Our results are $f_B=0.196(6)$ GeV, $f_{B_s}=0.236(7)$ GeV and $f_{B_s}/f_B =1.207(7)$, 
agreeing well with earlier results using the temporal axial current. 
By combining with these previous results, we provide updated values of $f_B=0.190(4)$ GeV, $f_{B_s}=0.229(5)$ GeV and $f_{B_s}/f_B = 1.206(5)$.  
\end{abstract}


\maketitle

\section{Introduction}
\label{sec:intro}
Hadronic weak decay matrix elements containing $b$-quarks 
that are calculated in lattice Quantum Chromodynamics (QCD) are critical to the flavour 
physics programme of overdetermining 
the Cabibbo-Kobayashi-Maskawa (CKM) matrix in order to find signs of new 
physics. 
The accuracy of the lattice QCD results often limits 
the accuracy with which the CKM matrix elements can be determined 
and with which the  
associated unitarity tests can be performed~\cite{pdg}. 
It is therefore important both to improve and to test the accuracy of 
the lattice QCD results. This includes determining the lattice QCD values 
using a variety of different formalisms for $b$ quarks and light quarks, in addition to using 
different methodologies within a given formalism. 

It is now becoming possible to study heavy quarks up to the mass of the bottom quark
using relativistic 
formalisms~\cite{bcHISQ, fBsHISQ}, but this is relatively expensive numerically.
Consequently, to date, the most extensive studies of heavy 
quarks in lattice QCD have been done with nonrelativistic formalisms, 
such as NRQCD~\cite{nrqcd} or the Fermilab formalism~\cite{fnal} and 
its variants~\cite{Christ:2006us}. 
Relativistic formalisms have the advantage of simple continuum-like current 
operators that couple to the $W$ boson that can be chosen to be absolutely 
normalised, for example through the existence of a partially conserved axial 
current (PCAC) relation~\cite{fdshort}. 
The main issue with these formalisms is then controlling discretisation errors~\cite{HISQ}.  
In nonrelativistic formalisms the numerical calculation itself is more 
tractable, along with the control of discretisation errors, 
but the current operators have a nonrelativistic expansion 
and must have their normalisation matched to that of the appropriate continuum operator.
The expansion and the normalisation are the main sources of systematic uncertainty in 
these lattice QCD results. 
The comparison of lattice QCD values derived using nonrelativistic and 
relativistic formalisms provides 
a test of systematic uncertainties (see, for example,~\cite{DowdallfB} and~\cite{fBsHISQ}). However it is 
also important to provide tests of systematic uncertainties within a given 
formalism using different methods. Here we provide such a test of the NRQCD 
approach by normalising new sets of current operators 
that have not been used in this formalism before, then comparing results for 
the decay constants and form factors obtained to the previous determinations. 

The archetypal heavy meson weak decay process is annihilation 
of a $B$ meson to $\tau \nu$. The hadronic parameter which 
controls the rate of this process is the $B$ meson decay constant, proportional 
to the matrix element to create a $B$ meson from the vacuum with 
the temporal axial current containing a heavy quark field and a light antiquark field. 
The most precise calculation to date of the $B$ meson decay 
constant, $f_B$, uses improved lattice NRQCD and Highly Improved Staggered light 
quarks on gluon field configurations that include $u$, $d$, $s$ and 
$c$ quarks in the sea with multiple values of the lattice spacing and 
a $u/d$ quark mass going down to the physical point~\cite{DowdallfB}. That calculation used 
lattice QCD perturbation theory~\cite{monahanmatch} to normalise the temporal axial current  
operator through $\mathcal{O}(\alpha_s)$, $\mathcal{O}(\alpha_s\Lambda_{\text{QCD}}/m_b)$
and $\mathcal{O}(\alpha_s a\Lambda_{\text{QCD}} )$ and obtained a final uncertainty of 
$2\%$, including uncertainties from current operator matching and missing higher 
order current operators. 

Decay constants can also be defined in continuum QCD from pseudoscalar 
current operators using the PCAC relation. This is typically the method 
of choice for lattice QCD calculations using relativistic formalisms 
where a lattice PCAC relation allows the pseudoscalar current to be 
absolutely normalised. This enables the $D$ and $D_s$ decay constants to 
be obtained with $0.5\%$ uncertainties using the HISQ formalism~\cite{fdshort,fdlong,milcfd}. Here we normalise the NRQCD-light pseudoscalar current 
through $\mathcal{O}(\alpha_s)$, $\mathcal{O}(\alpha_s\Lambda_{\text{QCD}}/m_b)$ 
and $\mathcal{O}(\alpha_s a\Lambda_{\text{QCD}})$ and obtain a value for $f_B$ with 
similar uncertainty to that determined from the temporal axial current, providing 
a test of the systematic errors. 

$B$ meson exclusive semileptonic processes are important for the 
determination of CKM matrix elements through the matching of experimental 
decay rates to theoretical 
expectations as a function of momentum transfer. Here the hadronic parameters 
that encapsulate the information needed on QCD effects are the form 
factors, calculable in lattice QCD. For the case in which both 
initial and final mesons are pseudoscalars (e.~g.~$B \rightarrow \pi \ell \nu$) 
there are two form factors, a vector 
form factor and a scalar form factor. It is the vector form factor that 
gives the decay rate in the light lepton mass limit, but both form factors 
appear in the lattice QCD determination of the matrix elements of 
the vector current. The form factors can be separated by comparing 
spatial and temporal vector current matrix elements but additional 
information can also be obtained by determining the scalar form 
factor directly from the scalar current. Indeed this method 
has been used for the accurate determination of $D$ and $K$ meson semileptonic 
form factors in lattice QCD using the HISQ formalism~\cite{Nadk,jonnadk,GamizKpi}. 
Here we compare results using NRQCD-light scalar currents 
to those obtained using vector currents for $B \rightarrow \pi \ell \nu$. 
We discuss how this 
method will be used in improved `second generation' $B$ meson 
semileptonic form factor calculations now underway.  

The paper is organised as follows: in section~\ref{sec:norm} we derive 
the normalisation of the NRQCD-light scalar and pseudoscalar current operators;
in section~\ref{sec:latt} we combine this with the lattice calculation 
of the matrix elements of different components of the current to 
give results for decay constants and form factors; section~\ref{sec:conclusions} 
gives our conclusions, including planned future work using these results. 

\section{Normalisation of lattice NRQCD current operators}
\label{sec:norm}

Here we discuss the normalisation of the lattice NRQCD-HISQ current operators 
when the light quark is taken to be massless and follow the methodology 
laid out in~\cite{colinjunko1,colinjunko2}, along with most of the notation. 
We will start with a discussion of the temporal 
axial current and show the modifications that need to be made to those results to yield the normalisation of the pseudoscalar current. Results for the temporal vector/scalar case are then identical because 
of the chiral symmetry of the HISQ action. 

The matrix element of the appropriate temporal axial current 
defined in continuum QCD 
between the vacuum and pseudoscalar meson, $H$, at rest yields the 
meson decay constant, $f_H$, via the relation 
\begin{equation}
\label{eq:fdef}
\langle 0 | A_0 | H \rangle = f_HM_H 
\end{equation}
where $M_H$ is the meson mass.
The continuum QCD current operator can be systematically expanded 
in terms of lattice NRQCD-HISQ current operators 
(whose matrix elements can be determined in a lattice QCD calculation) 
as 
\begin{equation}
\label{eq:matchdef}
A_0 = \sum_j C_{j,A_0}(\alpha_s,am_b) J^{(j)}_{A_0,lat} \, ,
\end{equation}
where increasing $j$ corresponds to operators that are higher order in a 
relativistic expansion. The $C_j$ are dimensionless coefficients that 
compensate for the different ultraviolet behaviour between the continuum 
and lattice regularisations of QCD and hence they can be calculated in 
perturbation theory as a power series in the strong coupling constant, $\alpha_s$. 
The coefficients of powers of $\alpha_s$ will depend on the bare heavy 
quark mass in lattice units, $am_b$, which is the parameter appearing in the lattice NRQCD action 
(we use $b$ rather than the generic label $h$ since this is almost always 
the $b$ quark). 
Here we work through $\mathcal{O}(\alpha_s)$ and include the three 
operators ($j=0,1,2$) that allow us to match the current through 
$\mathcal{O}(\alpha_s \Lambda_{\text{QCD}}/m_b)$ and $\mathcal{O}(\alpha_s a\Lambda_{\text{QCD}})$. 
The determination of the $C_j$ is done most conveniently by choosing to match 
matrix elements of the left- and right-hand sides
of eq.~(\ref{eq:matchdef}) for a heavy quark to light quark scattering process 
induced by the current. The procedure then~\cite{colinjunko1,colinjunko2} is to:
\begin{itemize}
\item calculate the amplitude 
for such a process through 
$\mathcal{O}(\alpha_s)$ in continuum QCD; 
\item expand this amplitude through first order in powers of $1/M$ 
where $M$ is the heavy quark pole mass; 
\item choose lattice NRQCD-HISQ operators that reproduce the terms 
in this expansion and calculate the one-loop mixing matrix of these 
operators in lattice QCD perturbation theory 
using the same infrared regulation procedure as used in 
the continuum.
Infrared divergences must cancel between the continuum and lattice calculations 
in the end, since the two only differ in ultraviolet physics. 
Note also that the mixing matrix should be calculated at a 
pole mass that matches that of the continuum calculation; 
\item invert this mixing matrix to determine the (finite) $C_j$ coefficients
that will give the correct linear combination of lattice currents 
to produce the same one-loop scattering amplitude as in continuum QCD. 
\end{itemize}

\begin{figure}[t]
  \centering
  \includegraphics[width=0.25\textwidth]{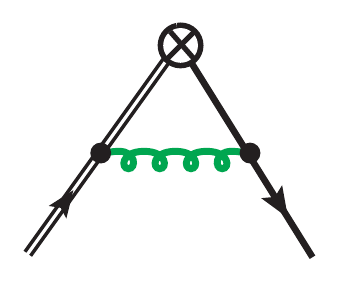}
  \caption{The Feynman diagram for the vertex renormalisation in continuum 
    QCD perturbation theory. The heavy quark is denoted by a double line, 
    the light quark by a single line and the exchange of a gluon by a curly line. 
    The current is denoted by a cross inside a circle. }
  \label{fig:vertex}
\end{figure}

The continuum calculation for the temporal axial current 
$\overline{q}(x)\gamma_5\gamma_0 h(x)$ was 
done in~\cite{colinjunko1} in the $\overline{MS}$ scheme using Feynman gauge, 
on-shell mass and wavefunction renormalisation 
and a gluon mass ($\lambda$) to regulate infrared divergences.  
$q(x)$ is the light quark field and $h(x)$ the heavy quark 
field satisfying the Dirac equation and the $\gamma$ 
matrices are the standard ones in Euclidean space-time. 
The key diagram to be calculated in 
continuum QCD is shown in Figure~\ref{fig:vertex}, where the double 
line represents an incoming heavy quark of momentum $p$, the 
single line an outgoing massless quark of momentum $p^{\prime}$ 
and the cross represents the current. The self-energy diagram must 
also be evaluated to determine the wave-function renormalisation. 
The result for the temporal 
axial current amplitude is given through $1/M$ as a combination of five matrix elements 
of Dirac spinors multiplied by factors of $p_0$, $p^{\prime}_0$ and 
$p\cdot p^{\prime}$ in~\cite{colinjunko1}. By using the Dirac equation for the light quark, and expanding the heavy quark energy and Dirac spinor to $1/M$, this is reduced to
\begin{equation}
\label{eq:A0exp}
\langle q(p^{\prime}) | A_0 | h(p) \rangle_{\mathrm{QCD}} = \eta^{(0)}_{A_0} \Omega^{(0)}_{A_0} + \eta^{(1)}_{A_0} \Omega^{(1)}_{A_0} + \eta^{(2)}_{A_0} \Omega^{(2)}_{A_0} . 
\end{equation}
\begin{table}[t]
  \caption{Values for the 3 $\zeta_{ij}$ one-loop mixing coeffficients 
    (defined in eq.~(\ref{eq:zij}))
    needed to determine the renormalisation of the lattice NRQCD-HISQ pseudoscalar/scalar 
    current from that of the temporal axial vector/temporal vector current 
    for massless HISQ quarks. Column 5 gives the one-loop NRQCD 
    mass renormalisation coefficient. These results were calculated and presented as the linear combination relevant for eq.~(\ref{eq:CA0res}) in \cite{monahanmatch} using the standard $v^4$-accurate NRQCD action (with stability parameter 
    $n=4$) and the individual values are given here. We also include new results 
    for a lighter $b$ quark mass, $am_b=1.22$, suitable for the 
    MILC `superfine' (0.06 fm) lattices. }
  \label{tab:zeta}
  \begin{center}
    \begin{tabular}{lllll}
      \hline
      \hline
      \vspace{-8pt}\\
      $am_b$ & $\zeta_{10}^{A_0}$  & $\zeta_{01}^{A_0}$ &  $\zeta_{12}^{A_0}$ & $Z_{m_b}$ \\ 
      \hline 
      3.297  &  -0.0958(1)  &  -0.1918(1)  &  0.029(4) & 0.167(1) \\
      3.263  &  -0.0966(1)  &  -0.1941(1)  &  0.030(4) & 0.176(1) \\
      3.250  &  -0.0970(1)  &  -0.1950(1)  &  0.031(4) & 0.178(1) \\
      2.688  &  -0.1144(1)  &  -0.2379(1)  &  0.060(4) & 0.262(1) \\
      2.660  &  -0.1156(1)  &  -0.2411(1)  &  0.060(4) & 0.264(1) \\
      2.650  &  -0.1157(1)  &  -0.2414(1)  &  0.061(4) & 0.267(1) \\
      2.620  &  -0.1171(1)  &  -0.2448(1)  &  0.062(4) & 0.272(1) \\
      1.910  &  -0.1539(1)  &  -0.3256(1)  &  0.093(4) & 0.434(1) \\
      1.890  &  -0.1553(1)  &  -0.3285(1)  &  0.095(4) & 0.448(1) \\
      1.832  &  -0.1593(2)  &  -0.3361(1)  &  0.097(4) & 0.466(1) \\
      1.826  &  -0.1595(2)  &  -0.3370(1)  &  0.098(4) & 0.468(1) \\
      1.220  &  -0.2258(5)  &  -0.4625(5)  &  0.116(5) & 0.714(1) \\
      \hline
      \hline
    \end{tabular}
  \end{center}
\end{table}
The coefficients are 
\begin{eqnarray}
\label{eq:Bexp}
\eta^{(0)}_{A_0} &=& 1 + \alpha_s B^{(0)}_{A_0} \nonumber \\
\eta^{(1)}_{A_0} &=& 1 + \alpha_s B^{(1)}_{A_0} \nonumber \\
\eta^{(2)}_{A_0} &=& \alpha_s B^{(2)}_{A_0} 
\end{eqnarray}
with
\begin{eqnarray}
\label{eq:Bres}
B^{(0)}_{A_0} &=& \frac{1}{3\pi}\left[ 3 \ln \frac{M}{\lambda} - \frac{3}{4} \right] \nonumber \\
B^{(1)}_{A_0} &=& \frac{1}{3\pi}\left[ 3 \ln \frac{M}{\lambda} - \frac{19}{4} \right] \nonumber \\
B^{(2)}_{A_0} &=& \frac{1}{3\pi}\left[ 12 - \frac{16\pi}{3}\frac{M}{\lambda} \right] .
\end{eqnarray}
The constituent matrix elements are
\begin{eqnarray}
\label{eq:omegadef}
\Omega^{(0)}_{A_0} &=& \overline{u}_q(p^{\prime})\gamma_5\gamma_0u_Q(p) \nonumber \\
\Omega^{(1)}_{A_0} &=& -i\overline{u}_q(p^{\prime})\gamma_5\gamma_0\frac{\bm{\gamma}\cdot\mathbf{p}}{2M}u_Q(p) \nonumber \\
\Omega^{(2)}_{A_0} &=& i\overline{u}_q(p^{\prime})\frac{\bm{\gamma}\cdot\mathbf{p^{\prime}}}{2M}\gamma_0\gamma_5\gamma_0u_Q(p) 
\end{eqnarray}
where $u_Q$ is a two-component spinor related to the Dirac spinor $u_h(p)$ (to $\mathcal{O}(1/M^2)$) by 
\begin{equation}
u_h(p) = \left[ 1 - \frac{i}{2M}\bm{\gamma}\cdot\mathbf{p}\right] u_Q(p)
\end{equation}
and where $u_Q$ satisfies $\gamma_0u_Q(p) = u_Q(p)$. 

We now carry out the continuum calculation to the same order for a pseudoscalar current 
$P = \overline{q}(x)\gamma_5h(x)$ and obtain
\begin{eqnarray}
\label{eq:Sexp1}
\langle q(p^{\prime}) | P | h(p) \rangle_{\mathrm{QCD}} &=& a_1 \overline{u}_q(p^{\prime})\gamma_5u_h(p) \nonumber \\
&+& a_2 \frac{\mathbf{p}\cdot\mathbf{p^{\prime}}}{M^2}\overline{u}_q(p^{\prime})\gamma_5u_h(p) 
\end{eqnarray}
where $\overline{u}_q$ and $u_h$ are Dirac spinors and 
\begin{eqnarray}
a_1 &=& 1 + \frac{\alpha_s}{3\pi}\left[\frac{13}{4}+3\ln\frac{\mu}{M}+3\ln\frac{\mu}{\lambda}\right] \nonumber \\
a_2 &=& \frac{\alpha_s}{3\pi} \left[ 4 - \frac{8\pi}{3}\frac{M}{\lambda} \right] ~.
\end{eqnarray}
Here $\mu$ is the scale parameter from dimensional regularisation and $\lambda$ is the gluon mass. 
We expand the heavy quark energy and Dirac spinor to $1/M$ and obtain
\begin{equation}
\label{eq:Sexp}
\langle q(p^{\prime}) | P | h(p) \rangle_{\mathrm{QCD}} = \eta^{(0)}_{P} \Omega^{(0)}_{P} + \eta^{(1)}_{P} \Omega^{(1)}_{P} + \eta^{(2)}_{P} \Omega^{(2)}_{P}  
\end{equation}
with $\eta^{(j)}_P$ defined in an analogous way to eq.~(\ref{eq:Bexp}) and 
\begin{eqnarray}
\label{eq:BPres}
B^{(0)}_P &=& B^{(0)}_{A_0} + \frac{1}{\pi}\left[ 2 \ln \frac{\mu}{M} + \frac{4}{3} \right] \nonumber \\
B^{(1)}_P &=& B^{(1)}_{A_0} + \frac{1}{\pi}\left[ 2 \ln \frac{\mu}{M} + \frac{8}{3} \right] \nonumber \\
B^{(2)}_P &=& B^{(2)}_{A_0} - \frac{4}{3\pi}.
\end{eqnarray}
\begin{table}[t]
  \caption{Values for the one-loop renormalisation factors 
    for the NRQCD-HISQ temporal axial vector current 
    (defined in eq.~(\ref{eq:A0z}))
    for massless HISQ quarks. The results for the temporal 
    vector current are identical. 
    These were calculated in~\cite{DowdallfB} from numbers in~\cite{monahanmatch}
    and are reproduced here. The results for $z_2^{A0}$ have changed 
    slightly for the heaviest masses because of an improved calculation of $\zeta_{02}$. }
  \label{tab:zA0}
  \begin{center}
    \begin{tabular}{llll}
      \hline
      \hline
      \vspace{-8pt}\\
      $am_b$ & $z_{0}^{A_0}$  & $z_{1}^{A_0}$ &  $z_{2}^{A_0}$ \\ 
      \hline 
      3.297  &  0.0238(20)  &  0.0242(28)  &  -1.014(6) \\
      3.263  &  0.0216(20)  &  0.0244(28)  &  -1.009(6) \\
      3.250  &  0.0220(10)  &  0.0240(22)  &  -0.999(6) \\
      2.688  &  0.0054(20)  &  0.0076(28)  &  -0.712(4) \\
      2.660  &  0.0056(20)  &  0.0074(28)  &  -0.698(4) \\
      2.650  &  0.0037(20)  &  0.0093(28)  &  -0.696(4) \\
      2.620  &  0.0011(20)  &  0.0069(28)  &  -0.690(4) \\
      1.910  &  -0.0071(20)  &  -0.0309(36)  &  -0.325(4) \\
      1.890  &  -0.0067(20)  &  -0.0313(36)  &  -0.318(4) \\
      1.832  &  -0.0027(20)  &  -0.0393(36)  &  -0.314(4) \\
      1.826  &  -0.0035(30)  &  -0.0395(42)  &  -0.311(4) \\
      1.220 &  0.0658(40)  & -0.0834(58) & 0.027(9) \\ 
      \hline
      \hline
    \end{tabular}
  \end{center}
\end{table}

Note that $B^{(0)}_P$ and $B^{(1)}_P$ have the same value ($(a_1-1)/\alpha_s$) coming from 
the first term on the right-hand side of eq.~(\ref{eq:Sexp1}).  
A check on these results comes from applying the continuum PCAC relation. 
This shows that the leading order term $B^{(0)}$ should differ between 
$P$ and $A_0$ by an amount that is the one-loop conversion factor between the 
pole and $\overline{MS}$ quark mass at scale $\mu$. 
Using $\gamma_0u_Q=u_Q$ the relationship between the operator matrix elements 
for the pseudoscalar and temporal axial current cases are
\begin{eqnarray}
\label{eq:omegaPdef}
\Omega^{(0)}_P &=& \overline{u}_q(p^{\prime})\gamma_5u_Q(p) = \Omega^{(0)}_{A_0} \nonumber \\
\Omega^{(1)}_P &=& -i\overline{u}_q(p^{\prime})\gamma_5\frac{\bm{\gamma}\cdot\mathbf{p}}{2M}u_Q(p) = -\Omega^{(1)}_{A_0} \nonumber \\
\Omega^{(2)}_P &=& i\overline{u}_q(p^{\prime})\frac{\bm{\gamma}\cdot\mathbf{p^{\prime}}}{2M}\gamma_0\gamma_5u_Q(p) = \Omega^{(2)}_{A_0} 
\end{eqnarray}
with a sign change for the leading relativistic correction, $j=1$. 
Exactly the same relations are obtained for the scalar case 
with respect to the temporal vector calculation. 
\begin{table}
  \caption{The results from this paper are the 
    values for the one-loop renormalisation factors 
    for the NRQCD-HISQ pseudoscalar current 
    (defined in eq.~(\ref{eq:Pz}))
    for massless HISQ quarks. 
    Results for the scalar current are identical.}
  \label{tab:zP}
  \begin{center}
    \begin{tabular}{llll}
      \hline
      \hline
      \vspace{-8pt}\\
      $am_b$ & $z_{0}^{P}$  & $z_{1}^{P}$ &  $z_{2}^{P}$ \\ 
      \hline 
      3.297  &  -0.0008(22)  &  0.2566(28)  &  -1.380(10) \\
      3.263  &  0.0044(22)  &  0.2538(28)  &  -1.373(10) \\
      3.250  &  0.0060(14)  &  0.2524(22)  &  -1.361(10) \\
      2.688  &  0.0386(22)  &  0.1850(28)  &  -1.016(9) \\
      2.660  &  0.0384(22)  &  0.1808(28)  &  -1.002(9) \\
      2.650  &  0.0393(22)  &  0.1823(28)  &  -0.998(9) \\
      2.620  &  0.0389(22)  &  0.1759(28)  &  -0.990(9) \\
      1.910  &  0.1191(22)  &  0.0501(36)  &  -0.563(9) \\
      1.890  &  0.1307(22)  &  0.0467(36)  &  -0.552(9) \\
      1.832  &  0.1447(22)  &  0.0315(36)  &  -0.544(9) \\
      1.826  &  0.1455(32)  &  0.0299(43)  &  -0.539(9) \\
      1.220  &  0.3278(40) & -0.1324(59) & -0.165(14) \\
      \hline
      \hline
    \end{tabular}
  \end{center}
\end{table}

The current operators needed in the lattice NRQCD calculation 
are readily identified from the $\Omega^{(j)}$ by replacing spinors 
with fields
and converting momentum factors to derivatives. 
This gives, for the temporal axial-vector case, 
\begin{eqnarray}
\label{eq:Jdef}
J^{(0)}_{A_0,lat} &=& \overline{q}(x)\gamma_5\gamma_0Q(x) \\ 
J^{(1)}_{A_0,lat} &=& -\frac{1}{2m_b}\overline{q}(x)\gamma_5\gamma_0\bm{\gamma}\cdot\overset{{}_{\shortrightarrow}}{\bm{\nabla}}Q(x) \nonumber \\ 
J^{(2)}_{A_0,lat} &=& -\frac{1}{2m_b}\overline{q}(x)\bm{\gamma}\cdot\overset{{}_{\shortleftarrow}}{\bm{\nabla}}\gamma_0\gamma_5\gamma_0Q(x)  \nonumber
\end{eqnarray}
where $m_b$ is the bare lattice NRQCD quark mass and $Q(x)$ is the two-component 
NRQCD field (i.e.~a four-component field with zero in the lower two-components). 
The analogous expressions for the pseudoscalar case mirror eq.~(\ref{eq:omegaPdef}).

The next step is to 
calculate the mixing matrix for the lattice operators in 
lattice QCD perturbation theory through one-loop. 
For the $A_0$ case 
\begin{equation}
\langle q(p^{\prime}) | J^{(j)}_{A_0,lat} | h(p) \rangle = \sum_j Z_{A_0,ij} \Omega^{(j)}_{A_0}
\end{equation}
with $Z_{A_0,ij}$ written as~\cite{colinjunko1,colinjunko2}
\begin{equation}
\label{eq:zij}
Z_{A_0,ij} = \delta_{ij} + \alpha_s\left\{\delta_{ij}\left[\frac{Z_q+Z_b}{2}+Z_{m_b}(1-\delta_{i0})\right]+\zeta^{A_0}_{ij}\right\}. 
\end{equation}
$Z_q$ is the coefficient of the one-loop term in the wavefunction renormalisation 
for massless lattice quarks, here in the HISQ formalism. 
Similarly, $Z_b$ is the coefficient of the one-loop term in the 
lattice NRQCD wavefunction renormalisation and 
$Z_{m_b}$ the coefficient of the one-loop mass renormalisation between the bare 
NRQCD quark mass and the pole mass~\cite{Dalgic}. 
This latter factor appears for $j=1,2$ because 
of the explicit mass factor in the operator and our choice to use the bare NRQCD 
mass in the NRQCD operators relevant for the lattice calculation. $\zeta_{ij}$ are the coefficients of the one-loop terms obtained from the renormalisation of the vertex diagram with $J^{(j)}$ 
at the vertex. Note that $Z_b$, $Z_{m_b}$ and $\zeta_{ij}$ are all functions 
of $am_b$ and must be evaluated at the value of $am_b$ being used in 
the lattice QCD calculation.    
\begin{figure}[t]
  \centering
  \includegraphics[width=0.47\textwidth]{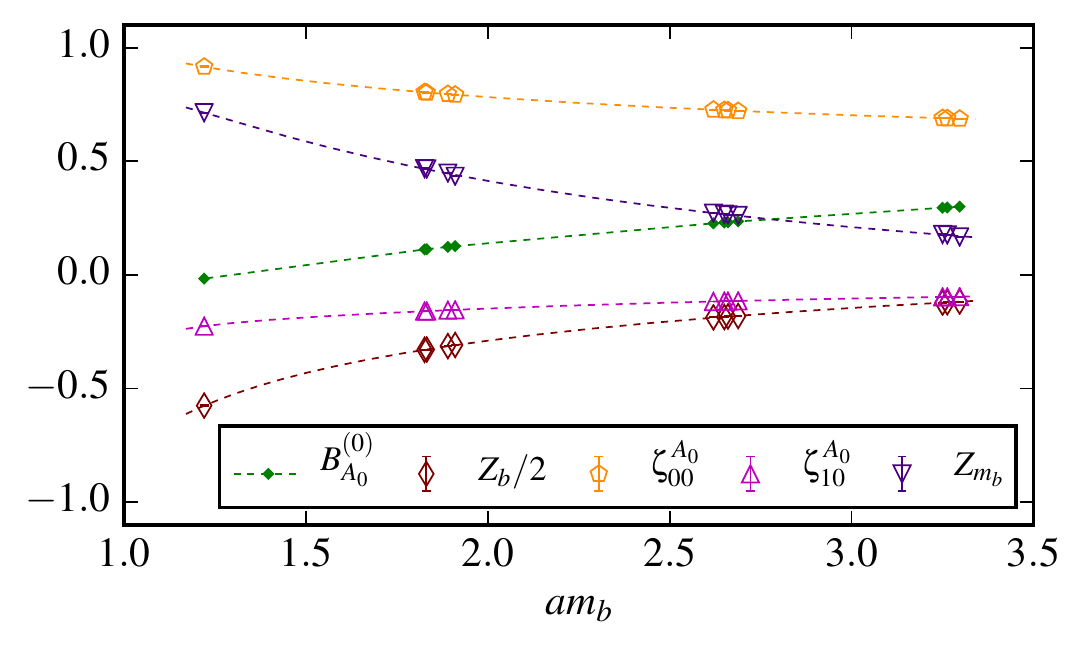}
  \caption{ The different contributions that make up the renormalisation 
    coefficients $z^{A_0}_0$ and $z^P_0$ as a function of the 
    bare heavy quark mass in the lattice NRQCD Hamiltonian, as given 
    in eqs.~(\ref{eq:CPres}) and~(\ref{eq:Pzval}). The infrared-finite 
    pieces are plotted for infrared-divergent contributions and $Z_q$ is not 
    included since it does not vary with $am_b$.}
  \label{fig:contributions}
\end{figure}

Peeling off the external states we can then write, using $A_0$ as an example, 
\begin{equation}
\label{eq:A0J}
A_0 = \sum_{i,j} \eta^{(i)}_{A_0} Z^{-1}_{A_0,ij} J^{(j)}_{A_0,lat}
\end{equation}
which determines the $C_j$ coefficients of eq.~(\ref{eq:matchdef}). 
To $\mathcal{O}(\alpha_s)$ 
\begin{equation}
\label{eq:zinvij}
Z^{-1}_{A_0,ij} = \delta_{ij} - \alpha_s\left\{\delta_{ij}\left[\frac{Z_q+Z_b}{2}+Z_{m_b}(1-\delta_{i0})\right]+\zeta^{A_0}_{ij}\right\} 
\end{equation}
so that, substituting in the results for the $\eta^{(i)}$ from eq.~(\ref{eq:Bexp}), 
we have~\cite{colinjunko1}
\begin{eqnarray}
\label{eq:CA0res}
C_{0,A_0} &=& 1 + \alpha_s\left(B^{(0)}_{A_0}-\frac{Z_q+Z_b}{2}-\zeta^{A_0}_{00}-\zeta^{A_0}_{10}\right) \nonumber \\
C_{1,A_0} &=& 1 + \alpha_s\left(B^{(1)}_{A_0}-\frac{Z_q+Z_b}{2}-Z_{m_b} -\zeta^{A_0}_{11}-\zeta^{A_0}_{01}\right) \nonumber \\
C_{2,A_0} &=& \alpha_s\left(B^{(2)}_{A_0}- \zeta^{A_0}_{02}-\zeta^{A_0}_{12}\right). 
\end{eqnarray}
$Z_q$ and $Z_b$ have logarithmic infrared divergences with $a\lambda$ as do 
$\zeta^{A_0}_{00}$ and $\zeta^{A_0}_{11}$. These cancel against the logarithmic 
divergences in 
$B^{(0)}_{A_0}$ and 
$B^{(1)}_{A_0}$ (see eq.~(\ref{eq:Bres}))
so that $C_{0,A_0}$ and 
$C_{1,A_0}$ are finite. Similarly the linear infrared divergence of $B^{(2)}_{A_0}$ 
is cancelled by a matching divergence in $\zeta^{A_0}_{02}$. 
Note that the explicit factors of $aM$ remaining in the $B^{(j)}_{A_0}$ 
will now be replaced by $am_b$, which is the same as $aM$ 
to this order in $\alpha_s$. 
Combinations of the $\zeta^{A_0}_{ij}$ that allow the $C_{j,A_0}$ to be 
determined are given for the NRQCD-HISQ 
case in~\cite{monahanmatch}. The calculation is done for the standard $v^4$-accurate 
NRQCD action~\cite{Lepage:1992tx, Gray:2005ur}, but the results are also correct for the $\alpha_s v^4$-improved 
NRQCD that we will use here~\cite{dowdallups}, because the impact of the $\alpha_s v^4$ improvement 
terms will only appear in the matching at $\alpha_s^2$.  

Here we recast the expansion of the QCD currents into a more natural combination 
of lattice QCD currents as 
\begin{eqnarray}
\label{eq:A0z}
A_0 &=& (1+\alpha_s z^{A_0}_0)\times \\
&&\left(J^{(0)}_{A_0,lat}+(1+\alpha_s z^{A_0}_1)J^{(1)}_{A_0,lat} + \alpha_s z^{A_0}_2J^{(2)}_{A_0,lat}\right) \nonumber
\end{eqnarray}
where, to $\mathcal{O}(\alpha_s)$, $(1+\alpha_sz^{A_0}_0)=C_{0,A_0}$, $\alpha_sz^{A_0}_2 = C_{2,A_0}$ and 
$\alpha_s z^{A_0}_1 = C_{1,A_0}-C_{0,A_0}$. 
The values for $z^{A_0}_0$, $z^{A_0}_1$ and $z^{A_0}_2$ were given in~\cite{DowdallfB} and are reproduced 
here in Table~\ref{tab:zA0}. The values are the same for the temporal vector current from the 
chiral symmetry of the HISQ action.  

\begin{figure}[t]
  \centering
  \includegraphics[width=0.47\textwidth]{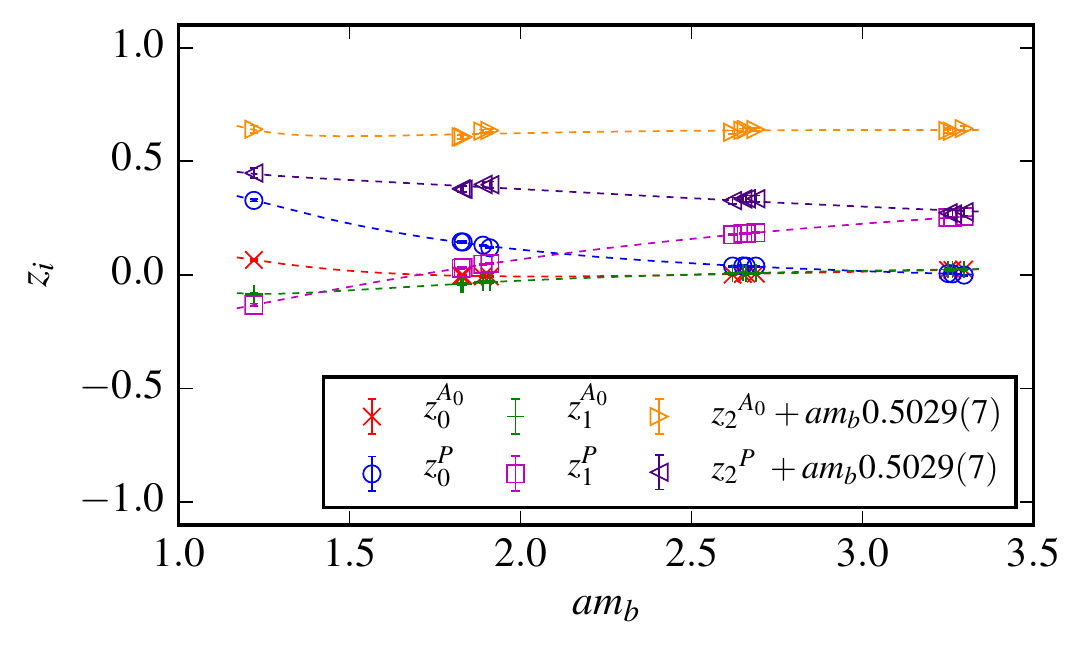}
  \caption{The $z_0$ and $z_1$ factors for the $\mathcal{O}(\alpha_s)$ matching of the 
    temporal axial and pseudoscalar NRQCD-HISQ currents to continuum QCD 
    (eqs.~(\ref{eq:A0z}) and~(\ref{eq:Pz})) plotted
    against the bare lattice $b$-quark mass. 
  }
  \label{fig:zs-massless}
\end{figure}

To perform the equivalent calculation for the pseudoscalar current we note
that the $\Omega^{(j)}_P$ are simply related to the 
$\Omega^{(j)}_{A_0}$ as in eq.~(\ref{eq:omegaPdef}) and so the $J^{(j)}_P$ are similarly related to $J^{(j)}_{A_0}$. 
Hence we do not need to perform a new calculation in lattice 
QCD perturbation theory. We simply need to reconstruct the 
mixing matrix for the pseudoscalar case from that of the 
temporal axial vector. 
We can then write
\begin{equation}
\label{eq:PJ}
P = \sum_{j} C_{j,P} J^{(j)}_{A_0,lat}
\end{equation}
and find
\begin{eqnarray}
\label{eq:CPres}
C_{0,P} &=& 1 + \alpha_s\left(B^{(0)}_{P}-\frac{Z_q+Z_b}{2}-\zeta^{A_0}_{00}+\zeta^{A_0}_{10}\right) \nonumber \\
C_{1,P} &=& -1 - \alpha_s\left(B^{(1)}_{P}-\frac{Z_q+Z_b}{2}-Z_{m_b} -\zeta^{A_0}_{11}+\zeta^{A_0}_{01}\right) \nonumber \\
C_{2,P} &=& \alpha_s\left(B^{(2)}_{P}- \zeta^{A_0}_{02}+\zeta^{A_0}_{12}\right) . 
\end{eqnarray}
Note the overall minus sign for $C_{1,P}$ as well as the fact that 
all of the $\zeta_{ij}$ factors with either $i$ or $j$ equal to 
1 now come in with opposite sign. These factors are all finite, so the 
$C_{j,P}$ are still manifestly infrared finite. 

Table~\ref{tab:zeta} gives results for the finite $\zeta$ factors, 
$\zeta^{A_0}_{10}$, $\zeta^{A_0}_{01}$ and $\zeta^{A_0}_{12}$, 
as well as $Z_{m_b}$ that allow us to determine the $C_{j,P}$ from the $C_{j,A_0}$ for a variety of values of the heavy quark mass in lattice units, $am_b$. These correspond to the values of $b$ quark masses used in our 
lattice NRQCD calculations that will be discussed in Section~\ref{sec:latt}. 

\begin{figure}[t]
    \centering
    \includegraphics[width=0.47\textwidth]{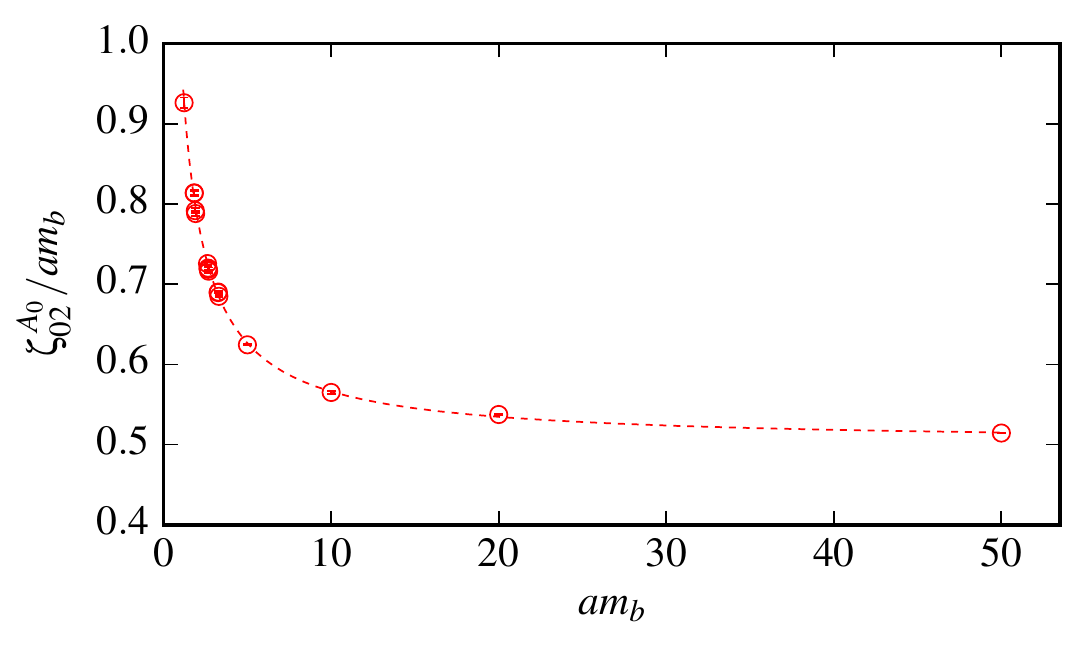}
    \caption{The one-loop mixing coefficient $\zeta^{A0}_{02}$ 
      for the temporal axial vector NRQCD-HISQ current for 
      massless HISQ quarks divided by the bare heavy quark 
      mass, $am_b$, and plotted against $am_b$. 
      This shows that $\zeta_{02}$ grows linearly with $am_b$ 
      as $am_b \rightarrow \infty$. 
    }
    \label{fig:zeta02-over-amb}
\end{figure}

For the pseudoscalar current case, we multiply both sides of eq.~(\ref{eq:PJ})
by the heavy quark mass in the $\overline{\mathrm{MS}}$ scheme 
at the scale $\mu$ and then, on the right-hand side, convert these into 
lattice NRQCD bare quark masses using the relation
\begin{equation}
\label{eq:massren}
m_b^{\overline{\mathrm{MS}}}(\mu) = m_b\left[1+\alpha_s\left(Z_{m_b}-\frac{2}{\pi}\ln\frac{\mu}{M}-\frac{4}{3\pi}\right)\right].
\end{equation} 
Values for $Z_{m_b}$ at a variety of $am_b$ values for lattice NRQCD are given in Table~\ref{tab:zeta}~\cite{monahanmatch}. 
Then we have 
\begin{eqnarray}
\label{eq:Pz}
P(\mu)m_b^{\overline{\mathrm{MS}}}(\mu) &=& m_b(1+\alpha_s z^P_0)\times \\
&&\hspace{-15mm}\left(J^{(0)}_{A_0,lat}-(1+\alpha_s z^P_1)J^{(1)}_{A_0,lat} + \alpha_s z^P_2J^{(2)}_{A_0,lat}\right). \nonumber
\end{eqnarray}
We find
\begin{eqnarray}
\label{eq:Pzval}
z^P_0 &=& z^{A_0}_0 + 2\zeta^{A_0}_{10} + Z_{m_b} \\
z^P_1 &=& z^{A_0}_1 + 2\zeta^{A_0}_{01} -2\zeta^{A_0}_{10} +\frac{4}{3\pi} \nonumber \\
z^P_2 &=& z^{A_0}_2 + 2\zeta^{A_0}_{12} -\frac{4}{3\pi}\, . \nonumber 
\end{eqnarray}
The values of appropriate $\zeta_{ij}$ and $Z_{m_b}$ 
given in Table~\ref{tab:zeta} enable us to determine the $z^P_j$ values from the 
$z^{A_0}_j$. The $z^P_j$ values are given in Table~\ref{tab:zP}. 
The values are the same for the scalar current due to the chiral 
symmetry of the HISQ action. 

Fig.~\ref{fig:contributions} shows 
the different contributions to $z^{A_0}_0$ and $z^P_0$ as a
function of $am_b$. This includes the finite pieces of each of the terms 
in eq.~(\ref{eq:CPres}) that vary with $am_b$ (thereby excluding $Z_q$). 
The different contributions are all of moderate size and show 
mild dependence on $am_b$ over the range of $am_b$ values 
that we use. 

In Fig.~\ref{fig:zs-massless} we plot $z_0$ and $z_1$ for 
the temporal axial vector and pseudoscalar cases. The magnitudes
of $z^{A_0}_0$ and $z^{A_0}_1$ are both very small and both 
have very little dependence on $am_b$, a fact previously 
remarked on in~\cite{DowdallfB}. $z^P_0$ and $z^P_1$ have larger 
magnitude and somewhat more dependence on $am_b$. However both are
still smaller than 1 across the range of $am_b$ values we use. 

Note that using $J^{(0)}$ alone in NRQCD to approximate 
either the temporal axial vector or pseudoscalar currents gives 
a larger renormalisation factor at $\mathcal{O}(\alpha_s)$. 
This is because the coefficient that represents the `mixing-down' of $J^{(1)}$ 
into $J^{(0)}$, $\zeta_{10}$, reduces the size of the one-loop 
renormalisation of the combined current in both cases. 
That $J^{(0)}+J^{(1)}$ is much closer to the continuum 
current than $J^{(0)}$ will be demonstrated in an order-by-order 
comparison of results in Section~\ref{sec:latt}.  

The coefficients $z^{A_0}_2$ and $z^P_2$ have stronger $am_b$ 
dependence dominated by that in the mixing coefficient $\zeta_{02}$.  
This grows linearly with $am_b$ at large values of $am_b$ so 
that, as $am_b \rightarrow \infty$, the contribution of $J^{(2)}$ 
becomes an $\alpha_sa\Lambda_{\text{QCD}} $ correction term. 
In Fig.~\ref{fig:zeta02-over-amb} we show $\zeta_{02}/am_b$ up 
to large values of $am_b$ (much above those that we use in practice) 
where this behaviour becomes clear. At the $am_b$ values that we 
use the $\alpha_s a\Lambda_{\text{QCD}}$ and $\alpha_s \Lambda_{\text{QCD}}/m_b$ 
behaviour is intertwined. 
Values of $z^{A0}_2$ and $z^{P}_2$ with the linear $am_b$ term 
removed are shown in Fig.~\ref{fig:zs-massless}. 

Lattice QCD results can be combined with 
eqs.~(\ref{eq:A0z}) and~(\ref{eq:Pz}) 
to determine the hadronic matrix elements 
of the temporal axial vector/vector and pseudoscalar/scalar currents 
up to systematic uncertainties 
coming from missing higher order radiative and 
relativistic corrections (which will differ between the currents). 
In the Section~\ref{sec:latt} we will compare results 
for hadronic decay constants obtained using 
temporal axial or pseudoscalar currents
and form factors from temporal vector and scalar 
currents. The extent to which they agree is a test 
of our systematic uncertainties. 

\begin{table*}[t]
  \caption{Sets of MILC configurations~\cite{milc1,milc2} used here with their (HISQ) sea quark masses, 
    $m_l$ (=$(m_u+m_d)/2$), $m_s$ and $m_c$ in lattice units. $\beta=10/g^2$ is the QCD gauge 
    coupling and the lattice spacing, $a$, is determined using 
    the $\Upsilon(2S-1S)$ splitting~\cite{dowdallups,DowdallfB}. 
    The lattice size is $L_s^3 \times L_t$. Each ensemble contains around 1000 configurations and we take 16 time sources per configuration to increase statistics. }
  \label{tab:params}
  \begin{center}
    \begin{tabular}{llllllllll}
      \hline
      \hline
      Set&$\beta$&$a$ (fm)&$am_{l}^{sea}$&$am_{s}^{sea}$&$am_c^{sea}$&$am_{s}^{val}$&$am_{b}^{val}$&$L_s/a$&$L_t/a$\\
      \hline
      1 & 5.80 & 0.1474(5)(14)(2) & 0.013 &0.065 & 0.838 & 0.0641 & 3.297 &16 & 48\\ 
      2 & 5.80 & 0.1463(3)(14)(2) & 0.0064 &0.064 & 0.828 & 0.0636 & 3.263 &24 & 48\\ 
      3 & 5.80 & 0.1450(3)(14)(2) & 0.00235 &0.0647 & 0.831 & 0.0628 & 3.25 &32 & 48\\ 
      \hline
      4 & 6.00 & 0.1219(2)(9)(2) & 0.0102 & 0.0509 & 0.635 & 0.0522 & 2.66 &24 & 64\\ 
      5 & 6.00 & 0.1195(3)(9)(2) & 0.00507 & 0.0507 & 0.628 & 0.0505 & 2.62 &32 & 64\\ 
      6 & 6.00 & 0.1189(2)(9)(2) & 0.00184 & 0.0507 & 0.628 & 0.0507 & 2.62 &48 & 64\\ 
      \hline
      7 & 6.30 & 0.0884(3)(5)(1) & 0.0074 & 0.0370 & 0.440 & 0.0364 & 1.91 &32 & 96\\ 
      8 & 6.30 & 0.0873(2)(5)(1) & 0.0012 & 0.0363 & 0.432 & 0.0360 & 1.89 &64 & 96\\ 
      \hline
      \hline
    \end{tabular}
  \end{center}
\end{table*}

\section{Lattice calculation and Results}
\label{sec:latt}

\subsection{Lattice configurations and simulation parameters}
\label{subsec:par}

The gluon field configurations used here are listed in 
Table~\ref{tab:params}. They are `second-generation' MILC configurations~\cite{milc1,milc2}
using a gluon action fully corrected through $\alpha_s a^2$~\cite{hartgluon} and 
HISQ quarks~\cite{HISQ} with $u$, $d$, $s$ and $c$ ($n_f=2+1+1$) flavors in the sea. 
They include multiple values of the lattice spacing and multiple values of 
the $u/d$ (taken to be degenerate) sea quark mass varying from one fifth 
of the $s$ quark mass down to the physical value. 
On these gluon field configurations the $B$ and $B_s$ decay constants were calculated 
in~\cite{DowdallfB} using a radiatively improved (through $\alpha_sv_b^4$) 
NRQCD action for the $b$ quark~\cite{hammant,dowdallups}, the HISQ action for the lighter quark and 
an NRQCD-HISQ temporal axial current matched to continuum QCD following 
the process described in Section~\ref{sec:norm}. 
Here we will compare results using the pseudoscalar current 
matched to the same level of accuracy. In a similar way, the systematic 
uncertainties in the semileptonic form factor for $B \rightarrow \pi$ obtained 
from (the traditional method of) using a vector current can be 
tested by employing a scalar current. 
Since we are largely re-using results from earlier papers~\cite{DowdallfB, Colquhoun:2015mfa} 
we do not repeat technical details, for example on the NRQCD Hamiltonian, 
but refer the reader to those papers for more detail.   

\subsection{$B$ and $B_s$ meson decay constants}
\label{subsec:fBBs}
Using the PCAC relation of continuum QCD we can 
determine the $B$ meson decay constant, $f_B$, from the matrix 
element of the temporal axial current between the vacuum and a $B$ meson (at rest) as 
\begin{equation}
\label{eq:fBA0}
\langle 0 | A_0 | B \rangle = f_B M_B
\end{equation}
or from the product of the pseudoscalar density and the quark masses as
\begin{equation}
\label{eq:fBP}
(m_b+m_l)\langle 0 | P | B \rangle = f_B M_B^2 .
\end{equation}
Here $M_B$ is the $B$ meson mass. 
In~\cite{DowdallfB} the temporal axial current 
relationship of eq.~(\ref{eq:fBA0}) was used.  
$A_0$ was constructed from the 
leading and next-to-leading NRQCD-HISQ currents in a non-relativistic 
expansion and was matched to continuum QCD according to eq.~(\ref{eq:A0z}).  
This involves writing $A_0$ in terms of the lattice 
currents, $J^{(0)}_{A_0,lat}$, $J^{(1)}_{A_0,lat}$ and 
$J^{(2)}_{A_0,lat}$. In~\cite{DowdallfB} the matrix elements 
of each current between the vacuum and a $B$ meson are 
determined in lattice QCD, so that the matrix element of 
$A_0$ in eq.~(\ref{eq:fBA0}) can be obtained to the 
specified level of accuracy (the matrix elements for 
$J^{(1)}_{A_0,lat}$ and $J^{(2)}_{A_0,lat}$ are the same 
for a meson at rest). 

In eq.~(\ref{eq:Pz}) we give an expansion to the 
same order for the combination of quark mass and 
pseudoscalar density, $m_bP$, in terms of the same 
NRQCD-HISQ currents multiplied by the bare NRQCD 
quark mass. Because the matrix elements for each 
of the lattice NRQCD-HISQ currents are  
given in~\cite{DowdallfB} we can reconstruct the matrix 
element of $m_bP$ required on the left-hand side of eq.~(\ref{eq:fBP})  
and so determine the decay constant in a different way. 
This decay constant should agree with that determined from the temporal 
axial current up to the uncertainties quoted. These are dominated 
by systematic errors from missing higher order matching terms 
and relativistic current corrections~\cite{DowdallfB}.  

\begin{figure}[t]
  \centering
  \includegraphics[width=0.47\textwidth]{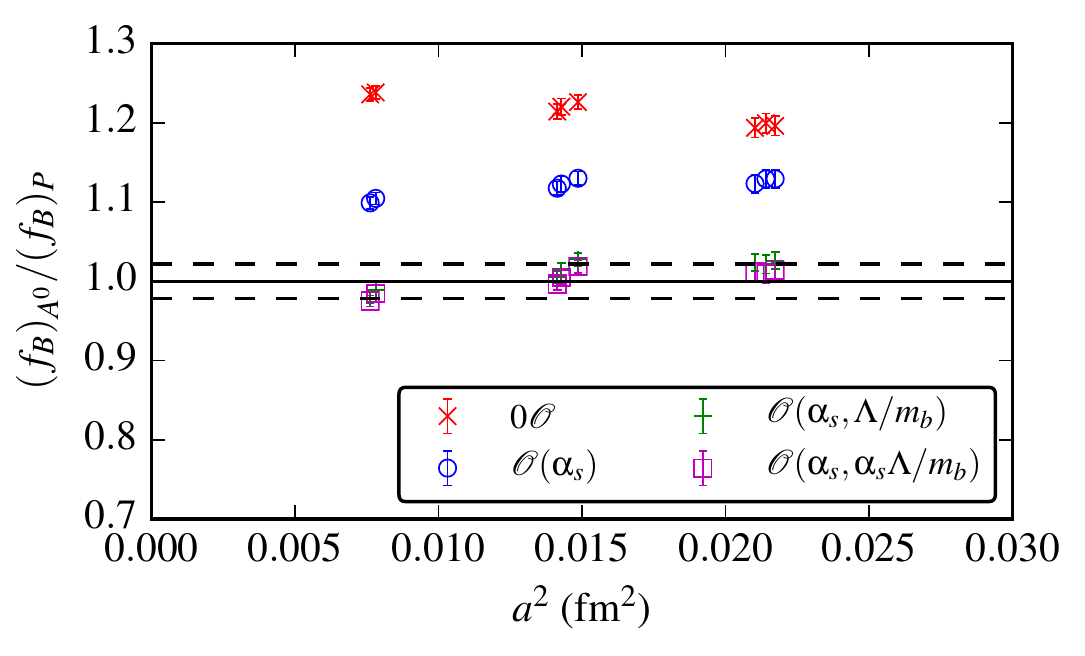}
  \includegraphics[width=0.47\textwidth]{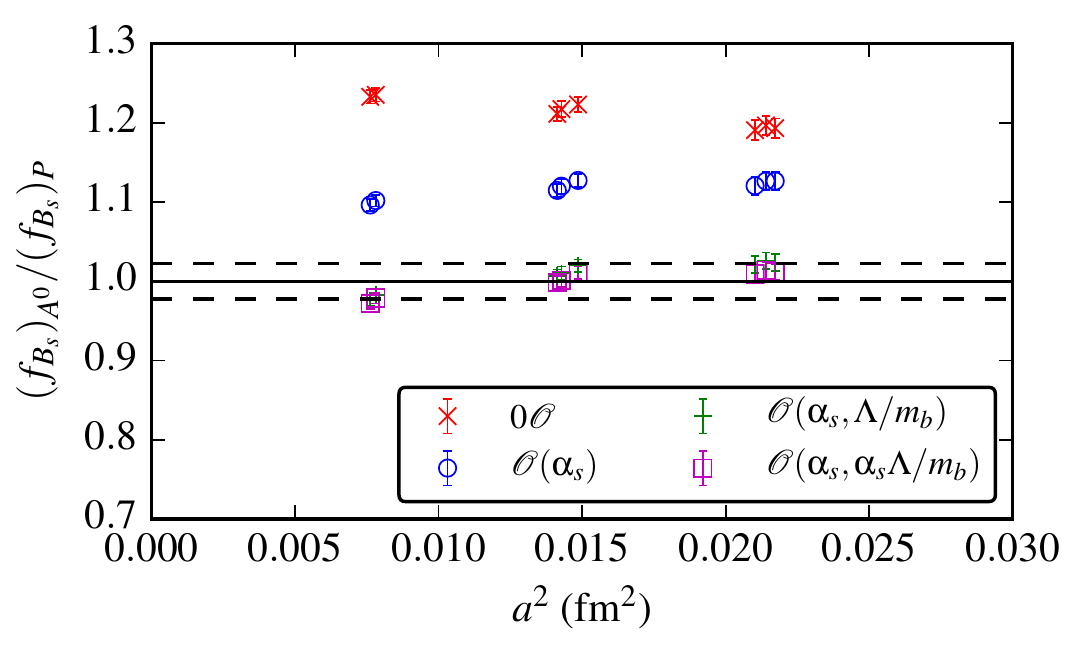}
  \caption{The ratio of the decay constant obtained using the 
    temporal axial current to that obtained using the 
    pseudoscalar density. Results are from 
    lattice QCD calculations of 
    the matrix element from a non-relativistic expansion of 
    the appropriate current operator between the vacuum and a $B$ meson (upper plot)
    or $B_s$ meson (lower plot). 
    Results from each of the ensembles of Table~\ref{tab:params} 
    are shown plotted against the square of the lattice spacing. 
    Red crosses denote the lowest (zeroth) 
    order result, while blue open circles include only $J^{(0)}_{A_0,lat}$
    but with $\mathcal{O}(\alpha_s)$ matching for that current. Green 
    pluses include $J^{(1)}_{A_0,lat}$ in a matching through $\mathcal{O}(\alpha_s)$ 
    and purple open squares include the full matching of eqs.~(\ref{eq:A0z})
    and~(\ref{eq:Pz}). The dashed lines show the relative uncertainty 
    on $f_B$ values quoted in ~\cite{DowdallfB}. 
  }
  \label{fig:fBPCAC}
\end{figure}

Figure~\ref{fig:fBPCAC} shows how well this process works 
order-by-order as relativistic current corrections and $\alpha_s$ 
matching terms are added in. The plot shows the ratio of 
the decay constant obtained using the temporal axial 
current to that using the pseudoscalar density. 
We use the results from~\cite{DowdallfB}, 
which gives matrix elements for each contribution to the current on each 
of the ensembles in Table~\ref{tab:params}. For each ensemble the bare 
NRQCD quark mass $am_b$ is tuned to that of the $b$-quark using the 
spin-average of $\Upsilon$ and $\eta_b$ masses and the $u/d$ quark 
mass, $am_l$, is given the value used for the light quark mass in the sea.  
The $s$ quark mass is tuned using a fictitious $s\overline{s}$ 
pseudoscalar meson whose properties are well-determined in lattice 
QCD~\cite{DowdallfKpi}. Values for the $\pi$, $K$ and $\eta_s$  meson masses made 
from these light quarks are given in~\cite{DowdallfKpi, DowdallHL}. 
We use $\alpha_s$ in the V-scheme at a scale $2/a$ in the operator matching as in~\cite{DowdallfB}. 

In the determination of $f_B$ from the pseudoscalar density in 
eq.~(\ref{eq:fBP}) there is a factor of $(1+m_l/m_b)$ on the left-hand side. 
We neglect this for the $u/d$ quark because, at the physical point, 
$m_l/m_b = 1/(52.55\times 27.4)$~\cite{mcnf4}. This is negligible 
compared to the other uncertainties. 
For the additional factor of 
$m_B$ on the right-hand side, which must be removed to take a ratio of 
the two different decay constants,  
we use the average of the charged and neutral experimental $B$ meson masses~\cite{pdg}. 
It has already been demonstrated that 
the lattice QCD result for the $B$ meson mass, using NRQCD for the 
$b$ quark, agrees with experiment at the 
physical value of the $u/d$ quark mass~\cite{DowdallHL}. 

In the upper plot of Figure~\ref{fig:fBPCAC} (for the $B$ meson) 
the lowest (zeroth) order result includes 
only the $J^{(0)}_{A_0,lat}$ current at tree-level, whose matrix 
element cancels in the ratio, and so the 
result is simply $m_b/M_B$.   
Not surprisingly, substantial differences are seen between the results 
of $\mathcal{O}(20\%)$, being the size of the binding energy of a $B$ meson. 
A significant improvement is seen on including $\alpha_s$ radiative corrections 
to the normalisation of $J^{(0)}_{A_0,lat}$ in the open blue circles. 
Note that, as remarked in Section~\ref{sec:norm}, 
the renormalisation of $J^{(0)}_{A_0,lat}$ differs from that 
of $J^{(0)}_{A_0,lat}+J^{(1)}_{A_0,lat}$ because 
of `mixing-down' effects encapsulated in $\zeta_{10}^{A_0}$. 
The differences between the two decay constants are now $\mathcal{O}(10\%)$ 
reflecting missing relativistic corrections of $\mathcal{O}(\Lambda_{\text{QCD}}/m_b)$. 
The green pluses and purple open squares now successively include 
the effect of $J^{(1)}_{A_0,lat}$ at tree-level and then 
$\alpha_s$ corrections multiplying the matrix elements of both 
$J^{(1)}_{A_0,lat}$ and $J^{(2)}_{A_0,lat}$ (which are the same here).  
The green pluses and purple open squares are very close together, 
since the final $\alpha_s$ corrections have little impact. 

The final result, correct through $\alpha_s\Lambda_{\text{QCD}}/m_b$, denoted by 
the purple open squares in Figure~\ref{fig:fBPCAC} 
is close to the solid line at 1.0, which indicates the same result is obtained for the 
decay constant from both the temporal axial current and pseudoscalar density. 
More importantly the differences from the value 1.0 of this ratio lie within 
the dashed lines that correspond to the $2.2\%$ relative uncertainty 
quoted for $f_B$ from the temporal axial current in~\cite{DowdallfB}. 
This uncertainty 
was dominated by an estimate 
of the uncertainties expected from missing higher order relativistic
current corrections and $\alpha_s^2$ matching errors. 
Since the results from the temporal axial current and pseudoscalar 
density will have different uncertainties from the missing $\alpha_s^2$ matching, the 
result of Figure~\ref{fig:fBPCAC} is a demonstration that the estimates 
of these uncertainties are realistic. 

The lower plot of Figure~\ref{fig:fBPCAC} repeats 
this exercise for the decay constant of the 
$B_s$ meson, again using results from~\cite{DowdallfB}. 
In this case, using eq.~(\ref{eq:fBP}), we do not neglect the 
$s$ quark mass on the left-hand side. Instead we write
\begin{equation}
\label{eq:fBsP}
\left(1+\frac{m_s}{m_b}\right)m_b\langle 0 | P | B_s \rangle = f_{B_s} M_{B_s}^2 .
\end{equation}
and take $m_s/m_b = 1.0/52.55(55)$~\cite{mcnf4}. 
We use the experimental value of the $B_s$ meson mass 
to determine a ratio of the decay constants from 
temporal axial current and pseudoscalar density. 
The lower plot of Figure~\ref{fig:fBPCAC} has identical features to that of 
upper plot. 
This is not surprising because the 
relative effect of the matrix elements of the relativistic 
current corrections is the same for $u/d$ and $s$ quarks 
as can be seen from the results in~\cite{DowdallfB}. 

\begin{figure}[t]
  \centering
  \includegraphics[width=0.47\textwidth]{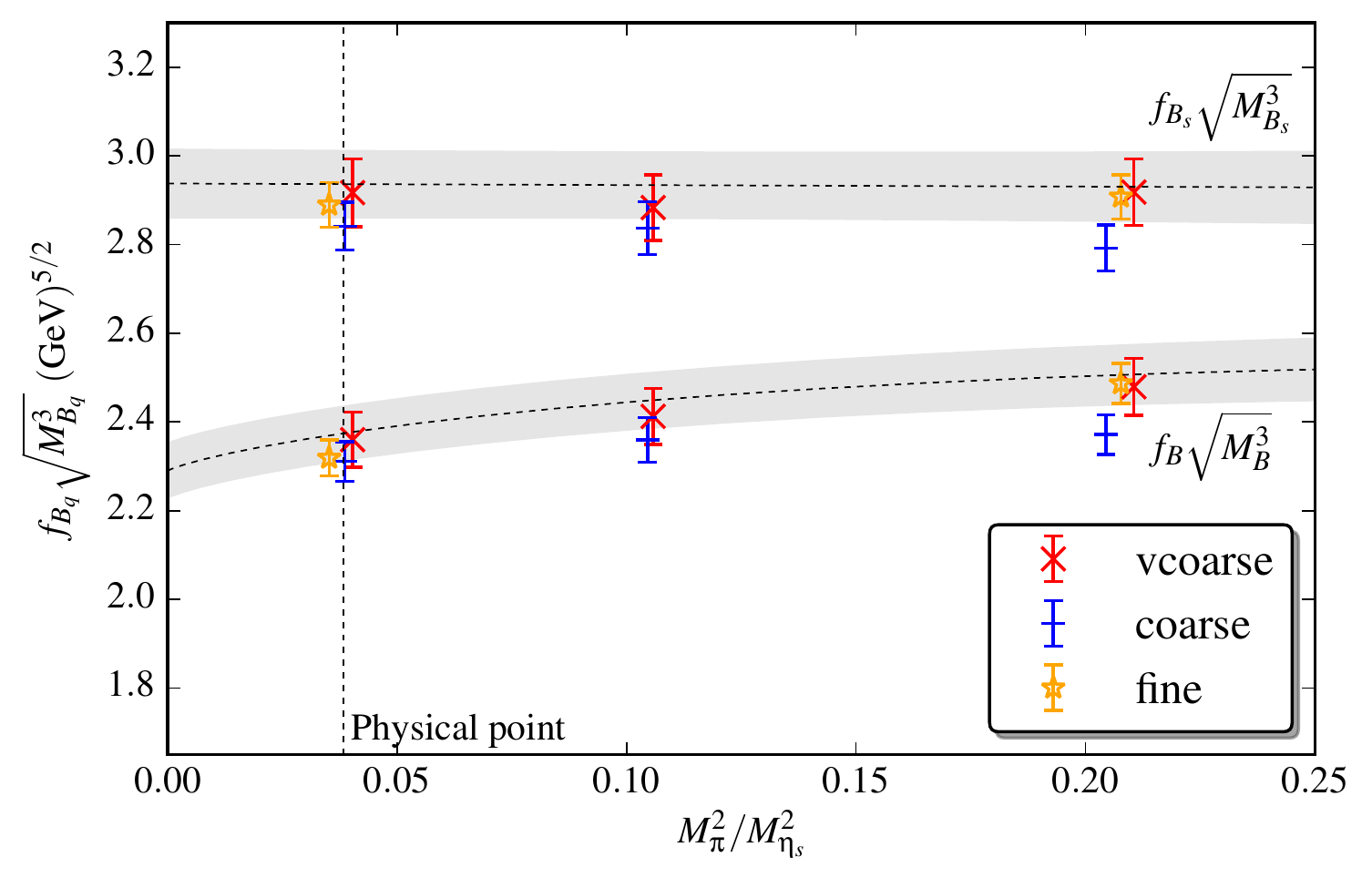}
  \caption{ Results for the decay constants of $B$ and $B_s$ 
mesons (multiplied by the $3/2$ power of the meson mass) 
for the ensembles in Table~\ref{tab:params}, obtained from the pseudoscalar 
current and plotted against 
the light quark mass in units of the strange quark mass (given 
as $M^2_{\pi}/M^2_{\eta_s}$). The grey bands show the results 
of the fit 
described in the text.  Errors on the data points include statistics/fitting 
only; the grey band includes the full error from the fit to lattice spacing and 
quark mass effects along with the perturbative matching uncertainty.  
  }
  \label{fig:fBfBs}
\end{figure}

\begin{figure}[t]
  \centering
  \includegraphics[width=0.47\textwidth]{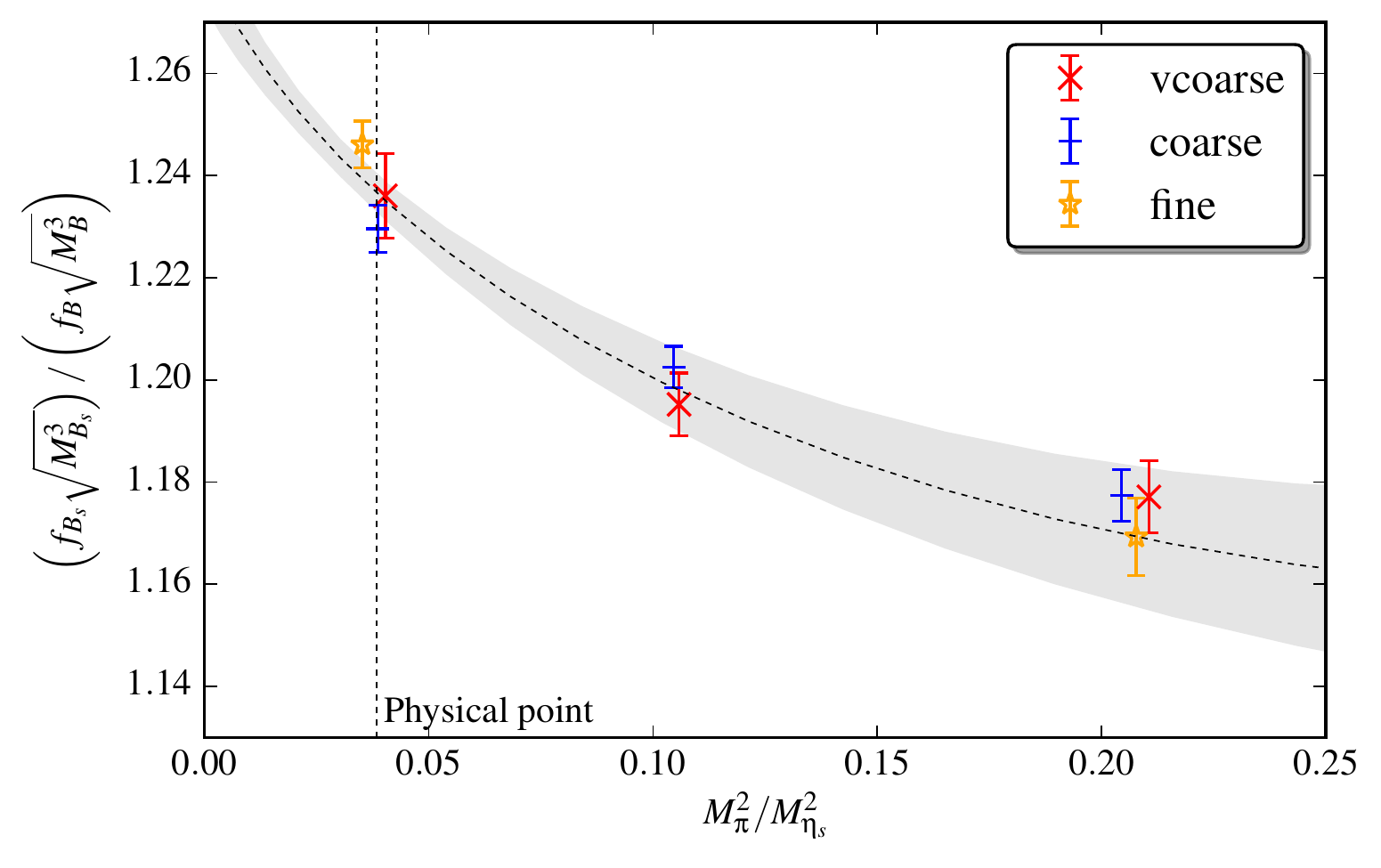}
  \caption{ Results for the ratio of $B_s$ to $B$ decay constants (multiplied by the 3/2 power of the ratio of meson mass) obtained from the 
pseudoscalar current. The data-points are as shown in Figure~\ref{fig:fBfBs}
and the grey band is the result of the fit described in the text, 
including uncertainties from lattice spacing and quark mass effects along with 
uncertainties from higher-order relatvistic corrections to the current. 
  }
  \label{fig:fBOfBs}
\end{figure}

We can complete the analysis by fitting the decay constant 
results for $f_B$ and $f_{B_s}$, obtained from the pseudoscalar 
density, as a function of $u/d$ quark mass and lattice spacing, 
to extract physical results for comparison to the 
final answers obtained using the temporal axial current. 
In the temporal axial current case, because of a normalisation factor 
from the meson states, the hadronic quantity naturally obtained from 
the lattice QCD calculation is $f_B\sqrt{M_B}$, denoted by $\Phi$. 
The contributions to $\Phi$ from $J^{(0)}_{A_0,lat}$ and 
$J^{(1)}_{A_0,lat}$ are tabulated in~\cite{DowdallfB}. 
The equivalent hadronic quantity corresponding to 
matrix elements of $m_bP$ is $f_B({M_B})^{3/2}$.
The values we need for $m_bP$ matrix elements are obtained by combining
the appropriate $\Phi$ values for $J^{(0)}$ and $J^{(1)}$ as in eq.~(\ref{eq:Pz}), 
including multiplcation by $am_b$ and then by an 
additional power of $1/a$ to 
convert to GeV units. In this case, the additional multiplication 
by $1/a$ will slightly increase the 
uncertainty in the values we are fitting because 
of the uncertainty in the determination of the lattice spacing. 

We plot $f_B(\sqrt{M_B})^3$ for the $B$ and $B_s$ mesons in Figure~\ref{fig:fBfBs}, 
as well as plotting our fits to the light quark mass and lattice spacing 
dependence that enables us to extract a physical result.  
Following~\cite{DowdallfB} we use a fit form
\begin{eqnarray}
f_{B}(\sqrt{M_{B}})^3(a,M_{\pi}) && = f_{B}(\sqrt{M_{B}})^3\times \\
&&(1 + d_1(\Lambda a)^2 + d_2(\Lambda a)^4) \times  \nonumber \\
&& \left(1 + b_{1,l} \frac{M_{\pi}^2}{\Lambda^2_{\chi}} - \frac{3(1+3g^2)}{4\Lambda^2_{\chi}}l(M_{\pi}^2)\right) \times  \nonumber \\
&& (1 + e_1\alpha_s^2[1+ e_2\delta m_b + e_3 \delta m_b^2]). \nonumber
\end{eqnarray}
Here $d_1$ and $d_2$ allow for discretisation effects and are 
also $\delta m_b$ dependent (suppressed for clarity). $b_1$ allows 
for dependence on the light quark mass, including the chiral logarithm, 
$l(M_{\pi}^2)$. For the $B_s$ case the chiral logarithm term is not present 
and $b_{1,l} \rightarrow b_{1,s}$. 
$e_1$ allows for $\alpha_s^2$ 
corrections from only matching to one-loop in perturbation theory, 
while $e_{2,3}$ allow for the fact that the higher order matching 
coefficients can in principle have $am_b$ dependence. 
These priors are given identical values as in~\cite{DowdallfB}, 
except for $e_1=0.0(3)$ as the pseudoscalar matching coefficients 
are slightly larger than their temporal axial-vector counterparts. 
Extrapolating to the physical point in the absence of 
electromagnetism, i.e.~$M_{\pi} = M_{\pi_0}$, where $m_l = (m_u + m_d )/2$, we 
find $f_B({M_B})^{3/2} = 2.37(7)$ GeV$^{\frac{5}{2}}$, 
$f_{B_s}({M_{B_s}})^{3/2}=2.94(8)$ GeV$^{\frac{5}{2}}$ 
and $f_{B_s}({M_{B_s}})^{3/2}/f_{B}({M_{B}})^{3/2}=1.237(7)$. 

\begin{table}[t]
  \caption{Full error budget for $f_{B_s}({M_{B_s}})^{3/2}$, $f_B({M_B})^{3/2}$ 
and their ratio as a percentage of the final answer.}
  \label{tab:Error}
  \begin{center}
    \begin{tabular}{llll}
      \hline
      \hline
      Error \% & Ratio & $f_{B_s}({{M_{B_s}}})^{3/2}$ & $f_{B}({M_{B}})^{3/2}$ \\\hline
      $a$ dependence: & $0.0$  & $1.1$ & $1.1$ \\      
      chiral:  & $ 0.02$ & $0.12$ & $0.13$ \\ 
      $g$: & $0.0$ & $0.01$ & $0.01$ \\                      
      stat/scale: & $0.3$ & $1.1$  & $1.1$ \\         
      operator: & $0.0$ & $2.0$ & $2.1$ \\            
      relativistic: & $0.5$ & $1.0$ & $1.0$ \\
      \hline
      total: & $0.6$ & $2.8$ & $2.8$ \\         
      \hline
      \hline
    \end{tabular}
  \end{center}
\end{table}

Our complete error budget is given in Table \ref{tab:Error}, with a breakdown that follows \cite{DowdallfB}. Errors arising from statistics, the lattice spacing, operator matching and chiral parameters are estimated directly from the fit. The remaining source of systematic error in the decay constants comes from missing higher order relativistic corrections to the current. As discussed in \cite{DowdallfB}, for the heavy-light system under consideration the higher order relativistic corrections will be of the size $(\Lambda_{\text{QCD}}/am_b)^2\simeq 0.01$, which we take for this component of the error.

We can convert the above results into values of the decay constants using the 
PDG masses \cite{pdg:2016} for $M_{B_l} = (M_{B_0}+M_{B^{\pm}})/2 =5.27963(15)$ GeV 
and $M_{B_s}=5.36689(19)$. 
Our final results for the decay constants obtained from the pseudoscalar current are 
$f_B=0.196(6)$ GeV, $f_{B_s}=0.236(7)$ GeV and $f_{B_s}/f_B = 1.207(7)$.

\begin{figure}[t]
  \centering
  \includegraphics[width=0.47\textwidth]{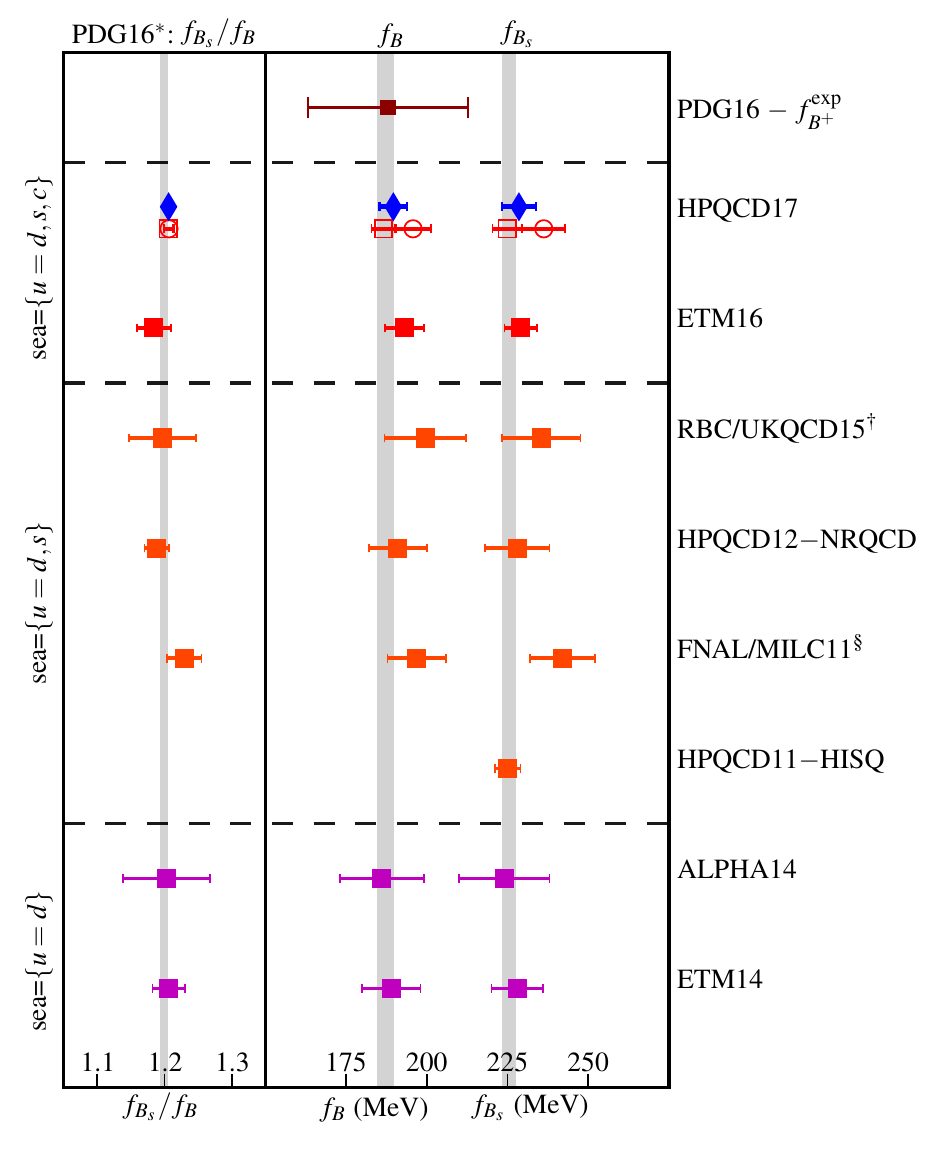}
  \caption{A summary of lattice QCD calculations for $f_B$, $f_{B_s}$ 
    and their ratio. The new results reported here are the those for the 
    pseudoscalar current given by the red open circles, showing good agreement 
    with earlier results (given by red open squares) using the temporal 
    axial current~\cite{DowdallfB}. 
    The average of these results is given by blue diamonds. Other results 
    in this summary are taken from~\cite{Bussone:2016iua, Christ:2014uea, Na:2012kp, Bazavov:2011aa, McNeile:2011ng, Bernardoni:2014fva, Carrasco:2013zta} and use 
    a variety of different quark formalisms for heavy and light quarks as 
    well as working with gluon field configurations that include different 
    numbers of flavours of sea quarks. The results for $f_B$ correspond to 
    those for a light valence quark of mass equal to the average of 
    $u$ and $d$ quark masses except for 
    `RBC/UKQCD15' which correspond to the neutral $b\overline{d}$ meson and 
    `FNAL/MILC11' which correspond to the charged $b\overline{u}$ meson. 
    The experimental result for the 
    charged $B$ meson is an average from the Particle Data Group~\cite{pdg}, 
    as are the grey bands. }
  \label{fig:worldav}
\end{figure}

As the same matrix elements are used as input when determining
the decay constant from temporal 
axial-vector and pseudoscalar currents, we must include correlations when 
performing an average of the results obtained here and in~\cite{DowdallfB}. 
As the different values for $f_{B_{(s)}}$ are slightly outside the $1\sigma$ error, 
we scale the error of the weighted averaged value by $\sqrt{\chi^2/\text{d.o.f.}}$ when 
the $\chi^2/\text{d.o.f.}\ge 1$. 
This is a conservative option and gives the scaled weighted averages as: 
$f_B=0.190(4)$ GeV, $f_{B_s}=0.229(5)$ GeV and $f_{B_s}/f_B = 1.206(5)$. 
These are shown, and compared to previous determinations, in Figure \ref{fig:worldav}, to be 
discussed further in Section~\ref{sec:conclusions}. 

\subsection{Scalar form factor for $B \rightarrow \pi$ decay}
\label{subsec:f0B}

\begin{figure}[t]
  \centering
  \includegraphics[width=0.47\textwidth]{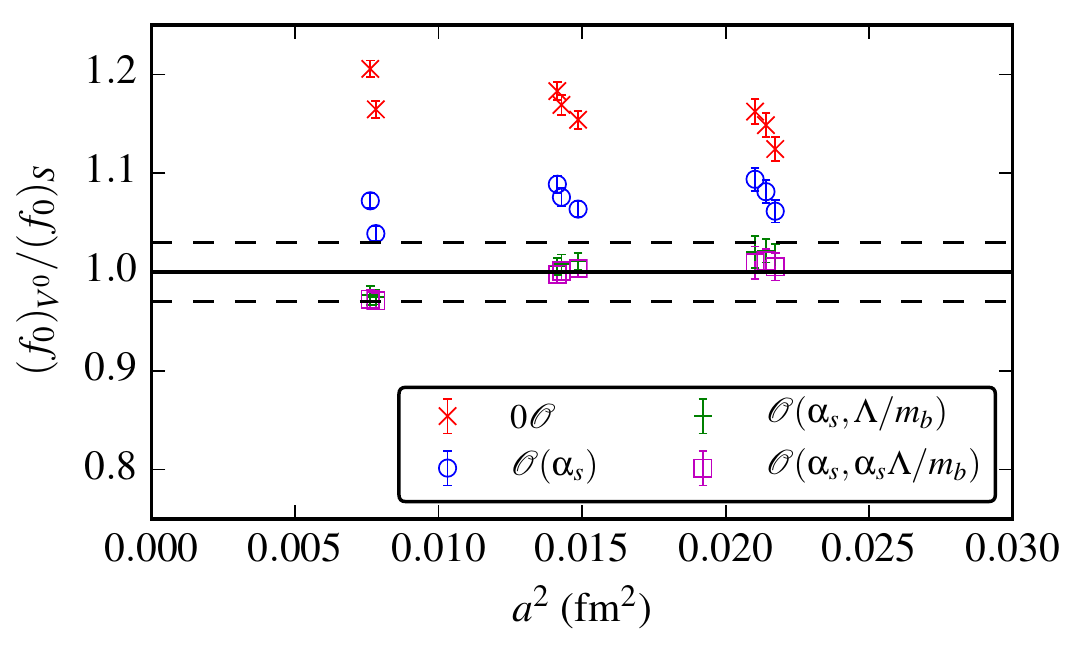}
  \caption{The ratio of the scalar form factor at zero recoil
    for the $B$ to $\pi$ decay obtained from using the temporal vector 
    current to that from using the scalar current. 
    Results are from 
    lattice QCD calculations of
    the matrix element of a non-relativistic expansion 
    of the appropriate current operator between a $\pi$ meson and a $B$ meson. 
    Results from each of the ensembles of Table~\ref{tab:params} 
    are shown plotted against the square of the lattice spacing. 
    Symbols are as in Figure~\ref{fig:fBPCAC}. 
    The dashed line gives the relative error on $f_0(q^2_{\mathrm{max}})$ 
    quoted in~\cite{Colquhoun:2015mfa}. 
  }
  \label{fig:f0PCVC}
\end{figure}

Another process where accurate determination of the matrix elements 
of heavy-light currents is required is for the weak semileptonic decays 
of $B$ mesons to light mesons. The archetypal process here is 
$B \rightarrow \pi \ell \nu$. The hadronic parameters needed to 
determine the rate for this decay are known as form factors and they 
are now functions of $q^2$, the squared 4-momentum transfer between the 
initial and final meson. For $B \rightarrow \pi \ell \nu$ there are 
two form factors which we will denote $f_+$ and $f_0$, but the experimental 
rate is only sensitive to $f_+$ if the final state lepton is light. 
The form factors are related to the matrix elements of vector and 
scalar currents by 
\begin{eqnarray}
\label{eq:Vf}
\langle \pi | V^{\mu} | B \rangle &=& f_+(q^2)\left[p^{\mu}_B+p^{\mu}_{\pi} - \frac{M_B^2-M_{\pi}^2}{q^2}q^{\mu} \right] \nonumber \\
&+& f_0(q^2)\frac{M_B^2-M_{\pi}^2}{q^2}q^{\mu} 
\end{eqnarray}
and 
\begin{equation}
\label{eq:Sf} 
\langle \pi | S | B \rangle = f_0(q^2) \frac{M_B^2-M_{\pi}^2}{m_b-m_l} .
\end{equation}
There is a kinematic constraint that $f_+(0)=f_0(0)$. 

At the zero recoil, maximum $q^2$, point we can compare the matrix elements 
of the temporal vector and scalar currents directly. At that point, where 
$q^0 = M_B - M_{\pi}$, 
\begin{equation}
\label{eq:Vfmax}
\langle \pi | V^0 | B \rangle = f_0(q^2_{max})(M_B+M_{\pi})
\end{equation}
and 
\begin{equation}
\label{eq:Sfmax}
\frac{m_b-m_l}{M_B-M_{\pi}}\langle \pi | S | B \rangle = f_0(q^2_{max})(M_B+M_{\pi})
\end{equation}
For currents made of NRQCD $b$ quarks combined with HISQ light quarks, the 
chiral symmetry of the HISQ action guarantees that the nonrelativistic 
expansion of the 
temporal vector current has the same form as for the temporal axial vector 
current given in eq.~(\ref{eq:A0z}). Likewise $m_bS$ has the same expansion 
as $m_bP$ given in eq.~(\ref{eq:Pz}). As for the case of the decay constant 
discussed in Section~\ref{subsec:fBBs}, the currents that appear in 
each order of the nonrelativistic expansion are the same for $S$ and $V^0$.   
Thus we can construct the matrix element of $m_bS$ given the lattice matrix 
elements of the different current contributions for $V^0$. These are 
given for the zero recoil situation in~\cite{Colquhoun:2015mfa} for 
the same 2+1+1 gluon field configurations as used for the decay constants 
in Section~\ref{subsec:fBBs}, and listed in Table~\ref{tab:params}. 
\begin{figure}[t]
  \centering  
  \includegraphics[width=0.47\textwidth]{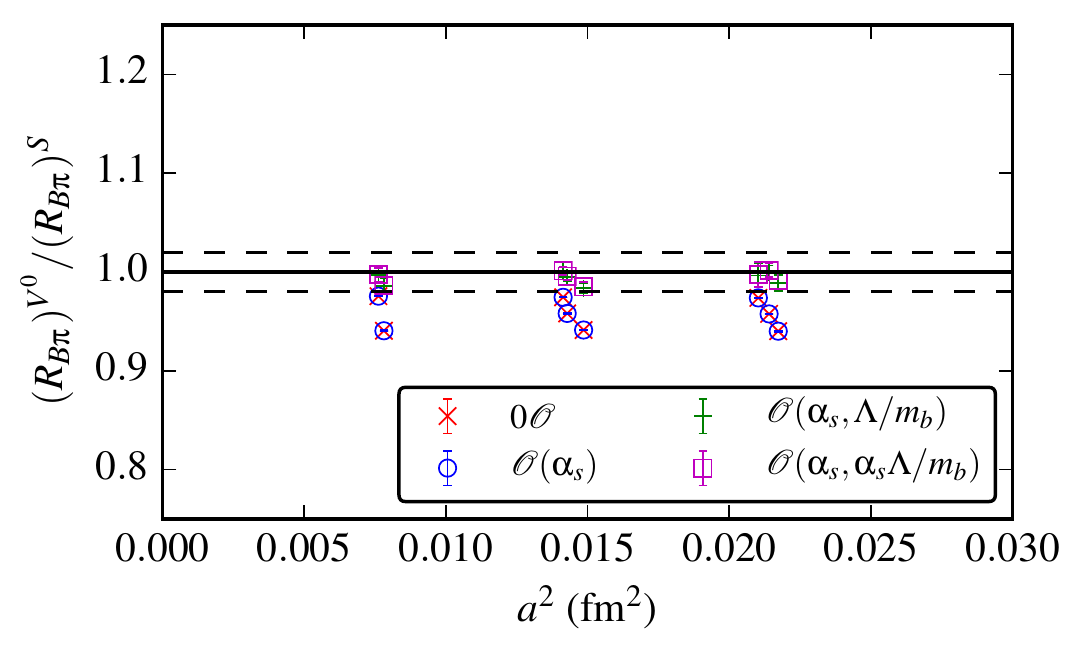}
  \caption{The ratio of $R_{B\pi}$ values (see text for definition) 
    obtained using a combination of temporal vector and temporal 
    axial vector currents to that from a combination of scalar 
    and pseudoscalar currents. 
    Results from each of the ensembles of Table~\ref{tab:params} 
    are shown plotted against the square of the lattice spacing. 
    Symbols are as in Figure~\ref{fig:fBPCAC}. 
    The dashed line gives the relative error on $R_{B\pi}$ from using the 
    temporal vector current for $f_0$ and temporal axial current for 
    $f_B$ quoted in~\cite{Colquhoun:2015mfa}. 
  }
  \label{fig:f0fBPCVC}
\end{figure}

Figure~\ref{fig:f0PCVC} shows the ratio of the scalar form factor 
at zero recoil determined from eqs.~(\ref{eq:Vfmax}) and~(\ref{eq:Sfmax})
using successively more accurate representations of the NRQCD-HISQ 
temporal vector and scalar currents from eqs.~(\ref{eq:A0z}) and~(\ref{eq:Pz}).
The matrix elements for the individual lattice current pieces are calculated 
in~\cite{Colquhoun:2015mfa}, noting that the matrix element of $J^{(2)}_{V^0}$
is equal to that of $J^{(1)}_{V^0}$ at zero recoil. In the additional mass factors 
on the left-hand side of eq.~(\ref{eq:Sfmax}) we ignore $m_l$ compared to 
$m_b$ as it is less than a $0.5\%$ effect across our range of light quark 
mass values. For $M_B$ we average the charged and neutral $B$ 
meson masses, as in Section~\ref{subsec:fBBs} and for $M_{\pi}$ we 
use the values appropriate to these ensembles given in~\cite{DowdallHL, DowdallfB}. 
 Figure~\ref{fig:f0PCVC} 
shows a very similar picture to that of Figure~\ref{fig:fBPCAC} with the 
ratio of the two results becoming closer to $1.0$ as non-relativistic current 
corrections are included and radiative corrections to them added in. 
With the most accurate matching that we have, the ratio of the results 
for the scalar form factor differs from one by less than the relative 
uncertainty on $f_0(q^2_{max})$ of 3\% quoted in~\cite{Colquhoun:2015mfa}. 
  
In~\cite{Colquhoun:2015mfa} the ratio of $f_0(q^2_{max})$ to the 
decay constant ratio $f_B/f_{\pi}$ was calculated to see if this ratio 
became one in the massless $\pi$ meson limit, as expected from soft 
pion theorems~\cite{Dominguez:1990mi, Wise:1992hn, Burdman:1992gh, Wolfenstein:1992xh}. This was indeed found to be the case, resolving a long-standing 
issue in the literature. The quantity calculated in~\cite{Colquhoun:2015mfa} 
was 
\begin{equation}
\label{rbpi}
R_{B\pi} = \frac{f_0(q^2_{max})(1+M_{\pi}/M_B)}{[f_B/f_{\pi}]}
\end{equation}
which becomes $f_0(q^2_{max})f_{\pi}/f_B$ as $M_{\pi} \rightarrow 0$. 
The piece of this ratio that involves $b$ quarks is $f_0(q^2_{max})/f_B$ 
and this was determined for NRQCD $b$ quarks from the ratio of 
the matrix element between $\pi$ and $B$ of the temporal vector current 
divided by the the matrix element between the vacuum and $B$ of the 
temporal axial vector current.  Because of the chiral symmetry of 
HISQ quarks the matching of these two NRQCD-HISQ currents to their 
continuum counterparts is the same. Then the overall renormalisation 
factor $(1+\alpha_sz_0 + \ldots)$ (see eq.~(\ref{eq:A0z})) cancels 
between them and the renormalisation uncertainty from missing 
$\alpha_s^2$ and higher order terms is much reduced. The ratio 
$R_{B\pi}$ can then be determined to high accuracy. 

Here we can also calculate $R_{B\pi}$, using the scalar 
current for $f_0(q^2_{max})$ and the pseudoscalar current for 
$f_B$. Again the overall renormalisation factor between the two 
will cancel (see eq.~(\ref{eq:Pz})). 
Figure~\ref{fig:f0fBPCVC} shows the ratio of $R_{B\pi}$ calculated 
in these two different ways as, once again, successively more 
accurate representations of the $b$-light currents are used. 
Now, because of cancellation of the overall renormalisation 
factors, there is no difference between 
the zeroth order result and the $\mathcal{O}(\alpha_s)$ result.  
Once $\alpha_s\Lambda_{\text{QCD}}/m_b$ corrections are included the 
ratio of $R_{B\pi}$ values is very close to one and well within 
the uncertainty of 2\% on $R_{B\pi}$ quoted in~\cite{Colquhoun:2015mfa}. 

\section{Conclusions}
\label{sec:conclusions}

Here we have given the matching calculation that enables matrix elements
of heavy-light scalar and pseudoscalar currents accurate through 
$\mathcal{O}(\alpha_s\Lambda_{\text{QCD}}/m_b)$ to be determined in lattice QCD using 
NRQCD $b$ quarks and HISQ light quarks. 
This expands the range of methods we can apply to $B$ physics using 
NRQCD and the tests we can do of our systematic error budget. 

In Section~\ref{subsec:fBBs} we determined the $B$ and $B_s$ meson decay 
constants using the pseudoscalar current and then compared to the 
previously determined values obtained from using the temporal axial current \cite{DowdallfB}. 
Since there is no PCAC relation connecting these two currents in 
lattice NRQCD, they do not have to give the same answer.
The way in which the nonrelativistic approximation to the continuum 
current is built up (albeit from the same ingredients) and the way in which it is 
renormalised are both different in the two cases. The 
systematic uncertainties from missing higher order terms will 
then also be different. We see in Figure~\ref{fig:fBPCAC} the results coming closer 
together as we include higher orders in the nonrelativistic 
expansion and matching on both sides. 
The final results, 
at the best accuracy that we can currently achieve, 
agree to within the expected remaining systematic uncertainty. 
This uncertainty is dominated by unknown $\alpha_s^2$ terms 
in the overall renormalisation (multiplying the leading current).  
The agreement is confirmation that the error budget is 
a reasonable one. 
Our results for the $B$ and $B_s$ decay constants from the pseudoscalar 
current are: 
\begin{eqnarray}
\label{eq:finalps}
f_B=0.196(6)\,\mathrm{GeV}, \\
f_{B_s}=0.236(7)\, \mathrm{GeV} \nonumber \\
f_{B_s}/f_B = 1.207(7)~. \nonumber
\end{eqnarray}

Our new results for the $B$ and $B_s$ decay constants (and 
their ratio) can be considered as independent values from 
those obtained using the temporal axial current in~\cite{DowdallfB}. 
They use the same raw lattice data in the form of matrix 
elements for the current components and so their statistical 
uncertainties are correlated. Their systematic uncertainties 
are not the same, however, since they come largely from unknown, and 
different, relativistic and $\alpha_s^2$ matching corrections. 
We can then perform a weighted average 
of the results arising from the two methodologies, 
including the statistical correlations, to obtain:  
\begin{eqnarray}
\label{eq:finalav}
f_B=0.190(4)\,\mathrm{GeV}, \\
f_{B_s}=0.229(5)\, \mathrm{GeV} \nonumber \\
f_{B_s}/f_B = 1.206(5)~. \nonumber
\end{eqnarray}
These results have very similar uncertainties to those in~\cite{DowdallfB} but 
do contain more information. 
 
Figure~\ref{fig:worldav} gives a summary of lattice QCD 
results for $f_B$, $f_{B_s}$ and their ratio. It includes 
results from a variety of light and heavy quark formalisms 
for calculations that have included at least 2 flavours of 
quarks in the sea. The most realistic version of QCD 
corresponds to the results in the top box, including the 
values we give here, where $u$, $d$, $s$ and $c$ sea quarks 
are incorporated. Our results have the additional 
advantage of including physical values 
for the $u/d$ sea quarks, taking $m_u=m_d$. The results for 
$f_B$ plotted in Figure~\ref{fig:worldav} correspond to 
a $B$ meson made with a light quark with the $u/d$ average mass. 
The grey bands show average values from~\cite{pdg}, where 
there is also discussion of the effects of isospin-breaking. 
The main message from Figure~\ref{fig:worldav} is that of
good agreement between the different lattice QCD results, 
which is another good test of systematic uncertainties. 

A similarly encouraging picture was given of the comparison 
of temporal vector and scalar current results for the scalar 
form factor for $B \rightarrow \pi$ decay in Section~\ref{subsec:f0B}.  
The calculations compared were done at zero recoil where the 
$\pi$ is at rest in the rest frame of the $B$. 
Here we discuss briefly the potential uses of our new method 
to determine the vector form factor 
$f_+$ away from the zero recoil point, where connection to experimental 
decay rates can be made for the determination of Cabibbo-Kobayashi-Maskawa 
matrix element $|V_{ub}|$. 

Power-counting in powers of the inverse heavy quark mass must be modified 
for current matrix elements away from the zero recoil point as the 
momentum of the light meson in the final state increases. Sub-leading currents 
that include a spatial derivative on the light quark field 
will have matrix 
elements that grow as $|\mathbf{p^{\prime}}|/m_b$ and these can become relatively 
large compared to the leading order current if $p^{\prime} > \Lambda_{\text{QCD}}$.  
The issue is discussed for the NRQCD-asqtad case in ~\cite{Dalgic:2006dt} 
where ratios of the sub-leading matrix elements to the leading matrix 
elements  between the $B$ meson and a heavy pion of the spatial vector 
current are shown. The matrix elements for currents denoted $J_k^{(2)}$ 
and $J_k^{(4)}$ grow as a proportion of the leading order ($J_k^{(0)}$) 
matrix element as the pion momentum is increased. This is also true, although 
not shown there, for the matrix elements of the temporal vector 
current $J_0^{(2)}$. These three currents are 
analogous to the current $J_{A_0,lat}^{(2)}$ (eq.~(\ref{eq:Jdef})) 
considered here, in having a 
derivative on the light quark field; for the spatial vector there are two 
such currents with different $\gamma$ matrix structures. 
Note that the matrix elements for the sub-leading currents that contain a 
derivative on the heavy-quark field show much more benign behaviour, as
might be expected.

The sub-leading currents with derivatives on the light quark field do 
not appear at tree-level in the expansion of the continuum heavy-light 
current (see eq.~(\ref{eq:A0z}) 
and so are suppressed by powers of $\alpha_s$. 
In the NRQCD-asqtad case the $\alpha_s$ coefficients of these sub-leading 
currents were calculated and turned out to be small for the largest 
contribution, from $J_k^{(4)}$~\cite{Dalgic:2006dt}. 
For the NRQCD-HISQ case these coefficients are only known for the  
temporal vector current (see Table~\ref{tab:zA0}~\cite{monahanmatch}).
As we move away from zero-recoil in $B\rightarrow \pi$ decay, 
systematic uncertainties from these sub-leading currents will grow if 
they are not included in our nonrelativistic expansion of the continuum 
current. It is therefore important to work with NRQCD-HISQ currents that 
do include the subleading currents with derivatives on the light quark field 
so that accuracy can be maintained as far from zero recoil as possible.  

Here we have provided a way of doing this by using the temporal 
vector and scalar currents (and eqs.~(\ref{eq:Vf}) and~(\ref{eq:Sf})), 
as used for example in calculations with 
purely HISQ quarks~\cite{Nadk, jonnadk}. As we have shown, both 
of these continuum currents can be written as a nonrelativistic expansion 
in NRQCD-HISQ currents that 
includes terms that will become $\mathcal{O}(\alpha_s|\mathbf{{p}^{\prime}}|/m_b)$ 
away from the zero recoil point. Uncertainties are then 
$\mathcal{O}(\alpha_s^2|\mathbf{{p}^{\prime}}|/m_b)$ and 
$\mathcal{O}(\alpha_s|\mathbf{{p}^{\prime}}|^2/m_b^2)$.  
This approach will be used in NRQCD-HISQ work on the second-generation 
2+1+1 HISQ configurations, extending~\cite{Colquhoun:2015mfa} away 
from zero-recoil. 

It should also be noted that the expansion for the pseudoscalar 
heavy-light current will allow more form factors to be separated out 
in the analysis of $B$ meson decays to light 
vectors~\cite{Horgan:2013pva, Horgan:2013hoa}. Processes such 
as $B_s \rightarrow \phi \ell^+\ell^-$ and $B \rightarrow K^* \ell^+\ell^-$
provide key opportunities for stringent tests of the 
Standard Model~\cite{Aaij:2013aln, Aaij:2014pli} and will need 
increasingly accurate lattice QCD results for comparison.

\subsection*{\bf{Acknowledgements}} 

We are grateful to the MILC collaboration for the use of their configurations and to 
R. Dowdall and B. Colquhoun for determination of the matrix elements needed in this calculation. 
Computing was done on the Darwin supercomputer at the University of 
Cambridge as part of STFC's DiRAC facility. 
We are grateful to the Darwin support staff for assistance. 
Funding for this work came from the 
Royal Society, the Wolfson Foundation and STFC. C.J.M. is supported in part by 
the U.S.~Department of Energy through Grant Number DE-FG02-00ER41132. This manuscript has been authored by Fermi Research Alliance, LLC under Contract No.~DE-AC02-07CH11359 with the U.~S.~Department of Energy, Office of Science, Office of High Energy Physics. The United States Government retains and the publisher, by accepting the article for publication, acknowledges that the United States Government retains a non-exclusive, paid-up, irrevocable, world-wide license to publish or reproduce the published form of this manuscript, or allow others to do so, for United States Government purposes.

\bibliography{hlnorm}

\end{document}